\documentclass[aps,prx,reprint,superscriptaddress,longbibliography,amsmath,amssymb]{revtex4-1}
\usepackage[utf8]{inputenc}
\usepackage{amsfonts}
\usepackage{physics}
\usepackage{graphicx}
\usepackage{overpic}
\usepackage{color}
\usepackage[dvipsnames]{xcolor}
\usepackage[hidelinks]{hyperref}
\usepackage{comment}

\usepackage[T1]{fontenc} 
\usepackage{enumitem}
\usepackage{booktabs}
\usepackage{float}

\usepackage{soul}

\usepackage{orcidlink}
\newcommand{\orcidalice}{\orcidlink{0000-0002-0563-5174}}
\newcommand{\orciddaniel}{\orcidlink{0000-0001-7658-3546}}
\newcommand{\orcidsimone}{\orcidlink{0000-0002-8882-2169}}
\newcommand{\orcidsebastian}{\orcidlink{0000-0001-9763-9131}}

\newcommand{\uulm}{Institute for Complex Quantum Systems, 
  Ulm University,
  Albert-Einstein-Allee 11, 89069 Ulm, Germany}
\newcommand{\unipd}{Dipartimento di Fisica e Astronomia "G. Galilei" \& Padua Quantum Technologies Research Center,
 Universit{\`a} degli Studi di Padova, Italy I-35131, Padova, Italy}

\newcommand{\pdinfn}{INFN, Sezione di Padova, via Marzolo 8, I-35131,
  Padova, Italy}
\newcommand{\ustuttgartIII}{Institute for Theoretical Physics III and
  Center for Integrated Quantum Science and Technology,
  University of Stuttgart, 70550 Stuttgart, Germany}

\newcommand{\rb}{r_{\rm \scriptscriptstyle B}}  
\newcommand{\rg}{r_{\rm \scriptscriptstyle G}}  
\newcommand{\al}{a}     
\newcommand{\rs}{r_{\rm \scriptscriptstyle S}}  
\newcommand{\ri}{r_{\rm \scriptscriptstyle H}}  
\newcommand{\doff}{d_{\rm \scriptscriptstyle offset}}  

\newcommand{\Hr}{H_{\rm \scriptscriptstyle Ryd}}  
\newcommand{\Ho}{H_{\rm \scriptscriptstyle OQS}}  

\newcommand{\Fdephasing}{\mathcal{F}_{\rms{D}}}

\newcommand{\TR}{T_{\rm \scriptscriptstyle R}}  
\newcommand{\PR}{P_{\rm \scriptscriptstyle R}}  

\newcommand{\rms}[1]{{\rm \scriptscriptstyle #1}}

\newcommand{\RotG}[3]{R_{\rm \scriptscriptstyle #1}^{\rm \scriptscriptstyle #2}\left(#3\right)}

\newcommand{\RGHZC}{RydberGHZ-C}

\newcommand{\Dmin}{D_{\rms{\min}}}
\newcommand{\Ds}{D(\rs)}
\newcommand{\Dczser}{D_{\rms{CZ-serial}}}
\newcommand{\Dser}{D_{\rms{serial}}}

\newcommand{\tlayer}{\tau_{\rms{Layer}}}

\newcommand{\figshare}{\tiny{}Image from \cite{Suppl}}
\newcommand{\figshares}{\tiny{}Images from \cite{Suppl}}
\newcommand{\figsharesapp}{\tiny{}Images from \cite{Suppl}}

\begin{document}

\title{Ab-initio tree-tensor-network digital twin for quantum computer benchmarking in 2D}

\author{Daniel Jaschke\orciddaniel${}^{*}$}
\affiliation{\uulm}
\affiliation{\unipd}
\affiliation{\pdinfn}
\email[]{Corresponding author: daniel-1.jaschke@uni-ulm.de}

\author{Alice Pagano\orcidalice}
\affiliation{\uulm}
\affiliation{\unipd}
\affiliation{\pdinfn}

\author{Sebastian Weber\orcidsebastian}
\affiliation{\ustuttgartIII}

\author{Simone Montangero\orcidsimone}
\affiliation{\uulm}
\affiliation{\unipd}
\affiliation{\pdinfn}


\begin{abstract}
  Large-scale numerical simulations of the Hamiltonian dynamics of a Noisy Intermediate Scale Quantum (NISQ) computer -- a digital twin -- could play a major role in developing efficient and scalable strategies for tuning quantum algorithms for specific hardware.
  Via a two-dimensional tensor network digital twin of a Rydberg atom quantum computer, we demonstrate the feasibility of such a program.  
  In particular, we quantify the effects of gate crosstalks induced by the
  van der Waals interaction between Rydberg atoms:
  according to an 8$\times$8 digital twin simulation based on the current state-of-the-art experimental setups, the initial state of a five-qubit repetition code can be prepared
  with a high fidelity, a first indicator for a compatibility with fault-tolerant
  quantum computing.
  The preparation of a 64-qubit Greenberger-Horne-Zeilinger (GHZ) state with about 700~gates yields a $99.9\%$
  fidelity in a closed system while achieving a speedup of $35\%$ via parallelization.
\end{abstract}

\maketitle

%


The present NISQ era of quantum computing poses extreme experimental, theoretical, and engineering challenges for all promising quantum computing platforms, being condensed matter or atomic-molecular-optical based ones~\cite{Almudever2017,Bruzewicz2019,Bertels2021,Pelofske2022}.
Indeed, identifying the best approaches, engineering solutions, and optimizing strategies at the physical, logical and algorithmic levels is necessary to maximize the capability of NISQ computers and unlock the fault-tolerant scalable era of general-purpose quantum computing~\cite{Preskill2018}. 
In the last two decades, these challenges have been mostly attacked at the level of few qubits, with impressive developments in, e.g., qubits and gate qualities~\cite{Cerfontaine2020,Zong2021,Koch2022}. However, the years ahead require achieving scalability and that will only be possible by understanding and characterizing the performances and limitations of the existing building blocks while functioning as one. For example, high-fidelity implementations will require taking into account also fast-decaying long-range interactions.
Moreover, to go beyond NISQ, decoherence effects shall be mitigated by reducing quantum circuits depth while quantum error-correcting codes will come at the price of additional gates: all this 
confront the software stack with further challenges, e.g.,
to what degree the gates on logical qubits can run in parallel. 

Here, we develop an efficient digital twin of a two-dimensional quantum processing unit (QPU) with access to a variety of compelling features, e.g., additional levels beyond the qubit states,
long-range interactions, and decoherence effects. These features of a large-scale digital twin
of the QPU will be fundamental to support the next decades of
developments, e.g., comparable to the impact that optimal control simulations had on the development of high-fidelity
single and two-qubit gates~\cite{Wilhelm2020,Mueller2022,Koch2022}.
Via tensor network methods~\cite{Haegeman2016,JaschkeOTN,Bauernfeind2020,Felser2021},
we perform two-dimensional large-scale classical simulations of a quantum computer running non-trivial quantum algorithms; tensor network methods allow one to overcome the curse of the exponentially increasing Hilbert space~\cite{Haner2017}.
We combine the
digital twin with a customized compiler and demonstrate how together they identify
 limiting factors of current or future hardware.
In this respect, the different topology and connectivity, e.g., 1-dimensional systems versus
2-dimensional ones, can lead to very distinct results in terms of the scaling of algorithms. 
We thus demonstrate how digital twins could guide the
development of future quantum algorithm compilers and transpilers~\cite{Haener2018,Heyfron2018,Murali2019}, 
specifically analyzing a Rydberg QPU in two dimensions.

Rydberg atoms trapped in optical tweezers represent one promising platform
for realizing a quantum computer~\cite{Saffman2010,Weimer2010,Loew2012,Saffman2016,Adams2019,Henry2021,Wu2021}. The Rydberg architecture has been impressively improved in various aspects over the last years and the execution of quantum algorithms of increasing complexity and circuit depth is in sight:  recent Rydberg experiments have demonstrated two-qubit gate fidelities beyond $98$\%~\cite{Graham2019,Madjarov2020,Madjarov2021c,Shi2022}.
A key ingredient to increasing the final computation quality and the achievable circuit complexity  
is evidently the ability to run gates in parallel. Parallel gate execution requires an independent parallel control and addressing of each qubit, which has been recently demonstrated~\cite{Labuhn2014,Browaeys2020,Bluvstein2021,Morgado2021,Ebadi2021,Scholl2021}.
Understanding if and how an algorithm can be parallelized
in the presence of long-range interactions, different connectivity, and spurious qubit crosstalks provides thus crucial insights to attack the next quantum computing engineering challenges.
The building blocks to experimentally implement the protocol suggested in the following in this manuscript are available, foremost
the CZ has been realized for example with $^{87}$Rb atoms~\cite{Levine2019}. 

To test the limits of the hardware on the digital twin, we analyze an algorithm that can
be easily verified via the measurement statistics in an experiment, can be scaled
with system size, is compatible with NISQ devices from the expected fidelity, and generates
an amount of entanglement compatible with simulations on a classical computer.
The preparation of large GHZ states fulfills these characteristics. Moreover, the GHZ state
is becoming a standard benchmark of the ability to control highly non-classical properties of quantum hardware. Since the seminal demonstration of 14-qubit GHZ states in a trapped ion quantum computer~\cite{Monz2011}, other platforms have accepted the challenge as well~\cite{Cruz2019, Omran2019, Mooney2021}. Recently, six-qubit GHZ states have been realized in a Rydberg quantum processor~\cite{Graham2022}. Therefore, this problem serves as an example of how
to use the digital twin from the problem statement to the statistics of projective measurements 
comparable to an actual experimental setup.
\begin{figure*}[t]
  \begin{center}
    \begin{minipage}{0.58\linewidth}
      \begin{minipage}{\linewidth}\begin{overpic}[width=1.0 \columnwidth,unit=1mm]{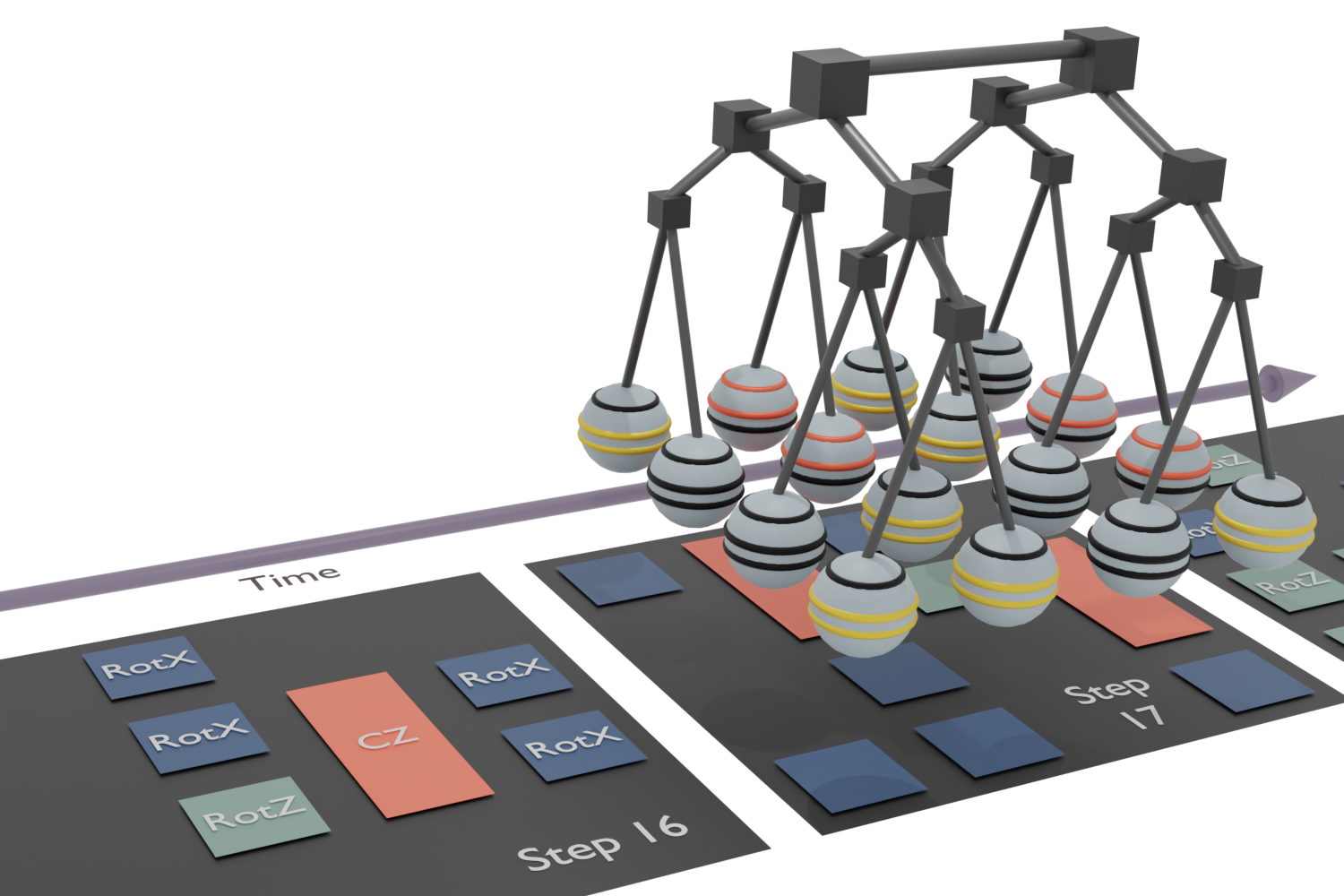}
        \put(0,67){a)}
        \put(0,38.2){\frame{{\includegraphics[scale=0.078]{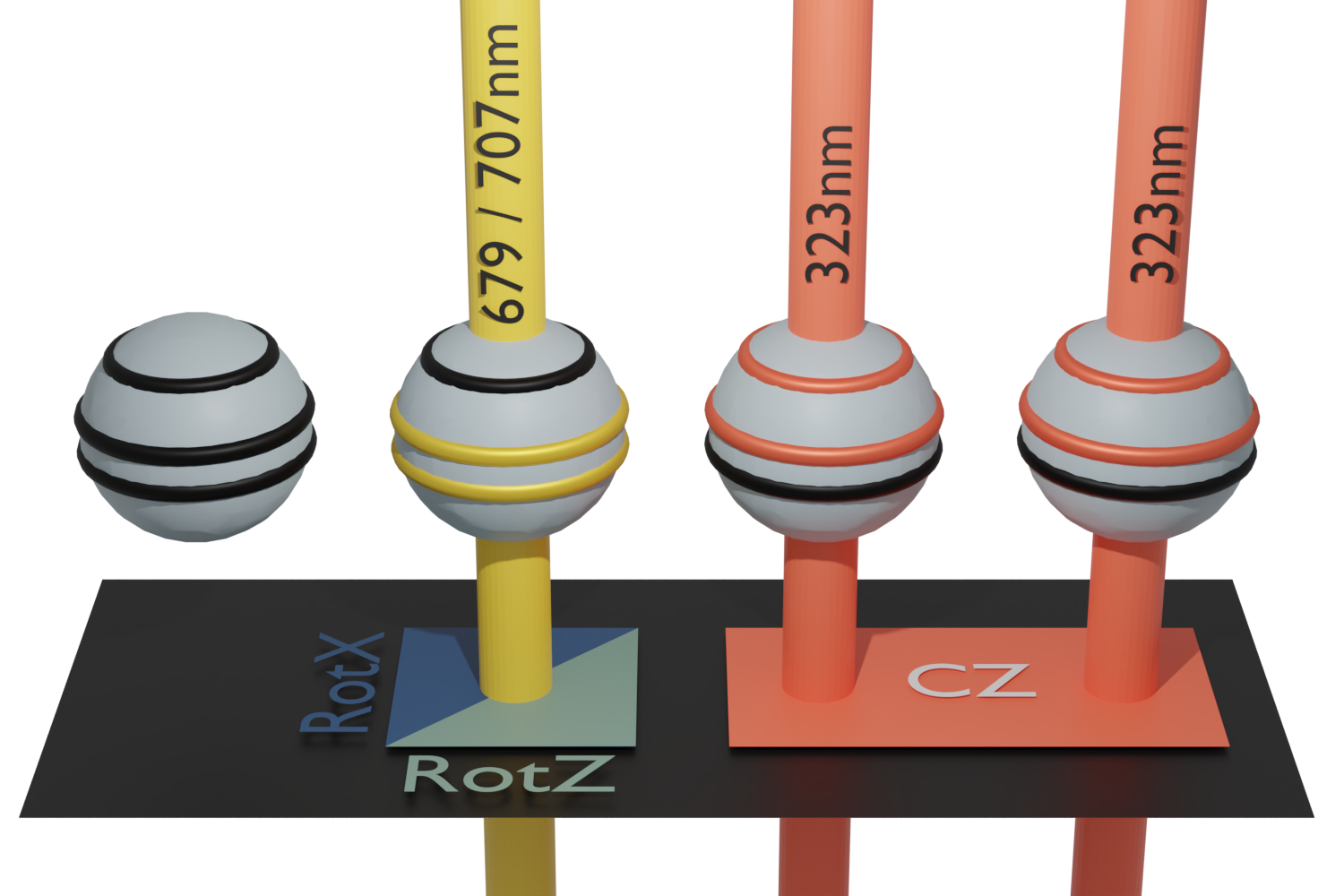}}}}
      \end{overpic}\vspace{0.1cm}\end{minipage}
      \begin{minipage}{1.0\linewidth}\raggedright b)
      \end{minipage}\hfill
      \begin{minipage}{\linewidth}\begin{overpic}[width=1.0 \columnwidth,unit=1mm]{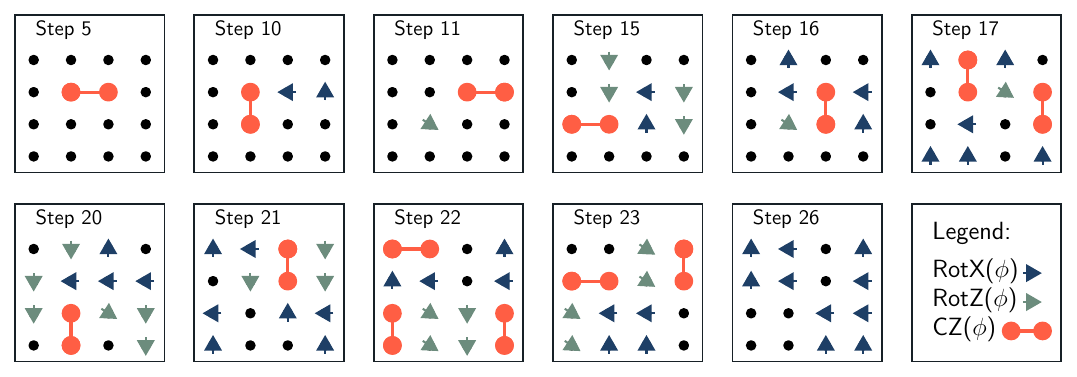}
      \end{overpic}\end{minipage}
    \end{minipage}\hfill
    \begin{minipage}{0.35\linewidth}
    \hspace{0.5cm}
       \begin{overpic}[width=0.515 \columnwidth,unit=1mm]{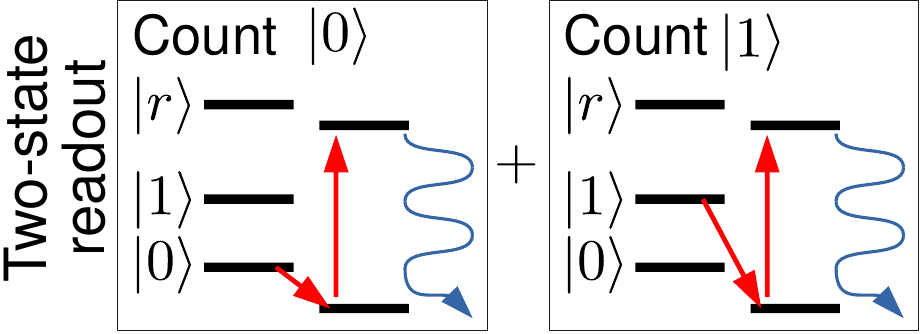}
       \end{overpic}
       \hspace{0.5cm}
       \begin{overpic}[width=0.275 \columnwidth,unit=1mm]{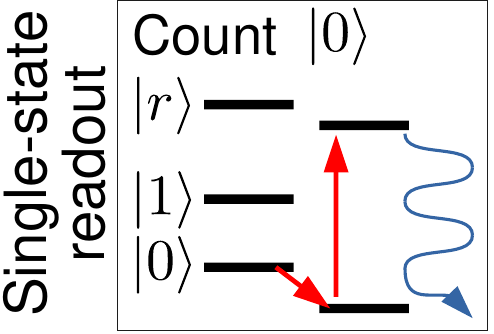}
       \end{overpic}
      \begin{minipage}{\linewidth}
      \begin{overpic}[width=1.0 \columnwidth,unit=1mm]{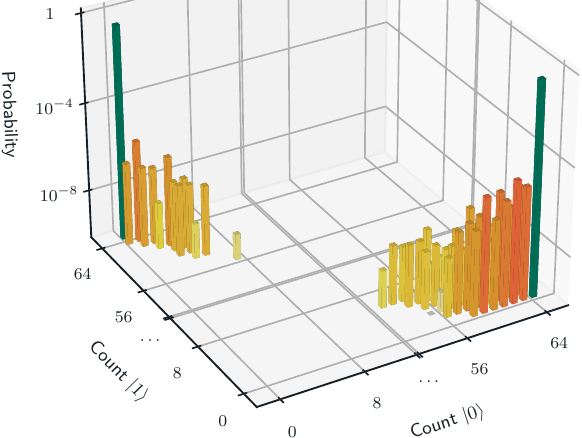}
        \put( -8,92){c)}
        \put(0, 92){\footnotesize{i)}}
        \put(65,92){\footnotesize{ii)}}
        \put(0, 70){\footnotesize{iii)}} 
        \put(40, 13){$\rg = \sqrt{10} \al$}
      \end{overpic}\vspace{0.2cm}\end{minipage}
      \begin{minipage}{\linewidth}
      \begin{overpic}[width=0.9 \columnwidth,unit=1mm]{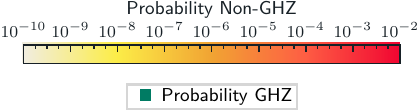}
      \end{overpic}\vspace{0.0cm}\end{minipage}
      \begin{minipage}{\linewidth}
      \begin{overpic}[width=1.0 \columnwidth,unit=1mm]{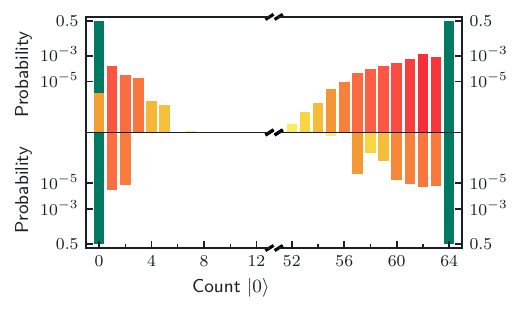}
        \put(35, 18){$\rg = \rs = \sqrt{16} \al$}
        \put(38, 47){$\rg = \sqrt{10} \al$}
        \put(72, 0){\figshares{}}
        \put(-3,54){\footnotesize{iv)}}
        \put(-3,32){\footnotesize{v)}} 
      \end{overpic}\end{minipage}
    \end{minipage}\hfill
    \caption{a)~\emph{Rydberg quantum computer setup:}
      We simulate a square grid of $^{88}\mathrm{Sr}$ Rydberg atoms trapped in optical
      tweezers. In each step, a layer of parallel gates is applied: 
      the controlled phase gates involve the strongly interacting Rydberg state $|r\rangle$, thus two of them applied simultaneously in close proximity introduce crosstalk errors. 
      A tree tensor network (TTN) simulates the Rydberg atoms
      as qutrits (states $\ket{0}$, $\ket{1}$, and $\ket{r}$).
      The lasers, individually addressing the atoms, implement the single-qubit and two-qubit gates (inset).
      b)~\emph{Parallelization:} Selected quantum algorithm layers of the GHZ state preparation
      in terms of selected gates native to the Rydberg platform. 
      Each square contains gates that are executed in parallel for a
      minimal distance between CZ gates $\rg \ge 2 \al$, where $\al$ is the lattice spacing.
      c)~\emph{Experimental measurement schemes:} Projective measurements for a perfect GHZ state yield a 50\% probability of
      counting either exactly 64 qubits in the $\ket{0}$ or $\ket{1}$ state for an
      8$\times$8 closed quantum system. We can choose between a two-state readout scheme
      in i) or a single-state readout scheme in ii), which both use an additional state.
      For $\rg = \sqrt{10}$, we compare the measurement statistics of readout scheme i) in iii) to readout scheme ii)
      in iv).
      We identify the states attributed to the GHZ state (in green) versus
      states introduced due to crosstalk (see color bar). As a consequence of a remaining
      population in the Rydberg state, the sum of qubits measured in $\ket{0}$
      and $\ket{1}$ does not necessarily add up to one,
      showing
      the effect of the remaining population in the Rydberg state $\ket{r}$. In comparison, $\rg = \sqrt{16} \al$
      limits the crosstalk to an acceptable amount as shown for the
      single-state readout scheme (see v) and is defined as the safe radius $\rs$
      to run gates in parallel.
                                                                                \label{fig:setup}}
  \end{center}
\end{figure*}
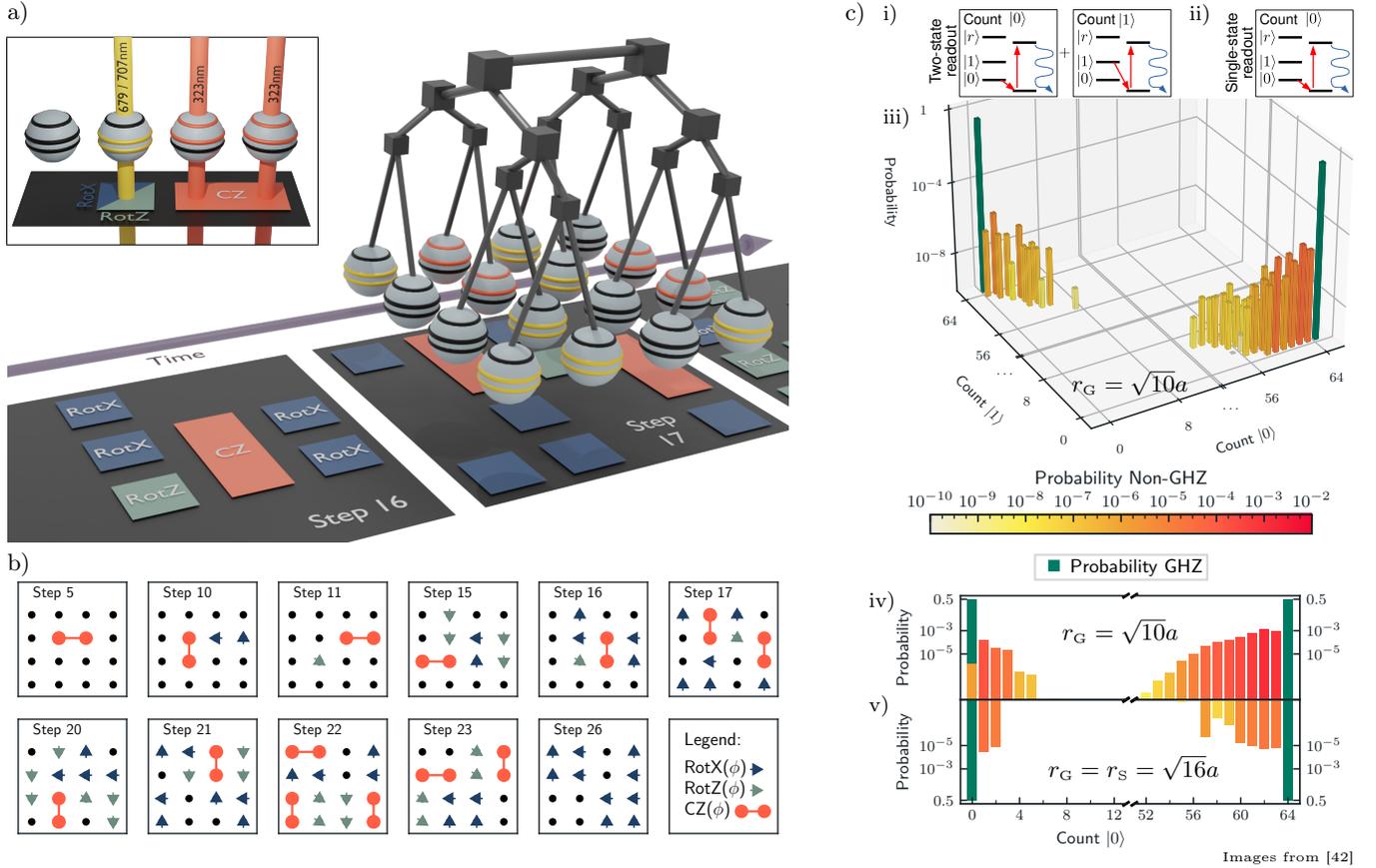

The digital twin allows one to establish the optimal radius beyond which gates can
run in parallel without significant crosstalk on the $^{88}$Sr Rydberg platform~\cite{Levine2019,Jandura2022,Pagano2022}.
In particular, we characterize the trade-off between higher parallelization
versus smaller errors due to crosstalk: on the one hand, a lower circuit depth
comes with higher error rates induced by the Rydberg interaction; on the other
hand, we pay for higher precision gates with larger circuit depths and an
increasing vulnerability to decoherence. As sketched in Fig.~1a), we
compile the circuit with a dedicated compiler \emph{\RGHZC{}} targeting the GHZ
state in a two-dimensional geometry. Then, we use realistic and conservative
parameters for the simulation of an $8\times 8$ strontium-88 setup~\cite{Pagano2022},
for example, taking into account long-range Rydberg interactions.
For a 64-qubit GHZ state, we obtain a state infidelity
of a $10^{-2}$ level for the closed quantum system and controllable crosstalk. The circuit depth
is only 15\% above the theoretical minimum for a 2D square system with
nearest-neighbor connectivity and the same gate set.
Finally, we demonstrate an application of the digital twin on parallel GHZ states generation, e.g.,
encountered in the initial preparation of quantum error-correcting five-qubit repetition code~\cite{Laflamme1996,Steane1996b,Gottesman2000}.

The numerical workhorse behind the ab-initio Hamiltonian-based emulation of a
parallel quantum computation is a tree tensor network (TTN) simulating a
square lattice of qutrits taking into account the states $\ket{0}$,
$\ket{1}$, and the Rydberg state $\ket{r}$~\cite{Silvi2017}. 
Within the family of tensor network algorithms~\cite{Schollwoeck2011,Orus2014,Montangero2018,Banuls2022},
the TTN is a powerful ansatz for two-dimensional systems.
The simulation of the system time evolution is implemented by exploiting recent progress in tensor network methods~\cite{Haegeman2016,JaschkeOTN,Bauernfeind2020,Felser2021}: in particular, the time-dependent variational
principle which supports the long-range
interactions required for picturing crosstalk and the scheduling of parallel
gates~\cite{Haegeman2016,Bauernfeind2020}. On the one hand, the combination of the many-body
simulations with optimal control results to include time-optimal gates and a compiler requires to master and merge 
all the existing state-of-the-art building blocks. On the other hand, exactly this
overarching approach distinguishes it from independently carried out analysis by the insight that can be gained: with an equivalent size of 101 qubits, the digital twins
sets the standard for emulating a QPU at the Hamiltonian level on a classical
computer.

The structure of the manuscript is the following: we focus on the prospect of running quantum algorithms
in parallel on the Rydberg platform by constructing a global GHZ state and
preparing multiple GHZ states on five qubits in Sec.~\ref{sec:parallel}. A detailed
description of the Rydberg system including open quantum system effects follows in Sec.~\ref{sec:rydberg}.
Afterward, we explain the technical aspects of the tensor network simulations in Sec.~\ref{sec:algo}
and the \RGHZC{} designed for the parallel GHZ preparation in Sec.~\ref{sec:rydberghz}.
We conclude with a brief summary and outlook.

\section{Parallel quantum algorithms}                              \label{sec:parallel}
%
The digital twin relies on an ab-initio Hamiltonian description of the platform. We consider that there is the potential to study all
qubit platforms in the way we demonstrate the simulations here for
Rydberg atoms.
The most challenging numerical aspects, i.e.,
the two-dimensional layout of superconducting hardware and the long-range
interactions of trapped ions, are combined in the Rydberg platform.
The two-dimensional structures in superconducting QPUs are continuously scaled up and
achieve 127 qubits~\cite{Niu2022}, which is already twice the size of the 64-qubit system studied
here. Although trapped ions systems are one-dimensional, their
strong long-range interactions allow an all-to-all connectivity~\cite{Brown2016}
which can lead to a rapid growth of entanglement and the classical resources needed
for the digital twin. In the following, we consider solely Rydberg
systems, i.e., neutral atoms.

For the Rydberg platform,
the physics required to understand the parallelization of the GHZ
state preparation can be summarized according to Fig.~\ref{fig:setup}.
We focus on the crosstalk in a closed quantum system in this section and
discuss open system effects in Sec.~\ref{sec:rydberg}.
Figure~\ref{fig:setup}a) shows a sketch of a
$4 \times 4$ setup of Rydberg atoms in optical tweezers; the lattice
constant of the grid  $\al$ introduces the first relevant length scale.
Within the Rydberg blockade radius $\rb$, only a single atom can be excited
to the Rydberg state due to the van der Waals interactions. We work at a
fixed Rydberg blockade radius $\rb = 4.98 \mu\mathrm{m} = 1.66 \al$ in
accordance with the experimental parameters proposed
in Ref.~\cite{Pagano2022}, which also provides the
time-optimal CZ gate used in the simulation here as well as the
parameters for a Strontium-88 setup; the pulse sequence for the
time-optimal CZ gate stems from optimal control.
%
On the one hand, the van der Waals interaction is giving rise to the Rydberg blockade which is exploited to implement an entangling gate between two atoms; on the other hand, the van der Waals interaction leads to possible crosstalk if multiple entangling gates act in parallel and atoms of two different CZ gates interact via van der Waals interactions. Consider as an example atoms A and B being the target of the first CZ gate and atoms C and D being the target of a second CZ gate; crosstalk arises if A or B interacts with atoms of the other gates, i.e., C and D, while both gates are driven at the same time. In contrast to the van der Waals interaction used to implement the CZ gate between A and B, additional atoms in the Rydberg state like C and D corrupt the pulse sequence and lead to what we refer to as crosstalk. 
Our compilation strategy
considers this condition and allows one to specify the minimal radius $\rg$ which
the \RGHZC{} and scheduler enforce between any two qubits
participating in different entangling gates executed in
parallel. We focus on the radius $\rg$ while the other length scales
are kept constant.
Finally, we are able to identify
the radius $\rs$ where the crosstalk is negligible in comparison to other errors and where the
algorithm is safely executed in parallel, i.e., we establish a criterion on the
fidelity of our state preparation. Figure~\ref{fig:setup}b) presents
the GHZ preparation for $\rg = 2 \al$ on a $4 \times 4$ square lattice
for a subset of native gates of the Rydberg platform which are implemented
for the digital twin. The radius $\rg$ changes
the circuit depth $D$ and the fidelity via crosstalk, while we keep the
lattice constant $\al$ and the Rydberg blockade radius $\rb$ fixed.
In the following paragraphs, we analyze these effects in detail. 

Figure~\ref{fig:setup}c) showcases a typical result enabled by the digital twin
simulation: the effects of crosstalk errors as they become visible
in projective measurements analog to an experimental setup. The fact
of simulating qutrits allows one to explore different readout schemes
either with a readout of both qubit states, see i) in Fig.~\ref{fig:setup}c),
or a single-state readout scheme where we choose the $\ket{0}$ state, see ii).
For each measurement of the $\ket{0}$ or $\ket{1}$ state, one needs
to transfer the corresponding state to an additional state for readout,
e.g., to the ground state of the optical qubit.
A single-state readout scheme, e.g., of state $\ket{0}$, avoids the
additional overhead of transferring the second state
$\ket{1}$ also to state for the readout and measuring it. The sum along the axis
\guillemotleft{}Count $\ket{1}$\guillemotright{}
of the histogram iii) for the two-state readout leads to the histogram
iv) of the single-state readout. The sum of the probabilities along one axis
underlines how the readout schemes differ in information:  counting
only the number of qubits in $\ket{0}$, the state $\ket{11 \ldots 11}$
originating in the GHZ state is indistinguishable from the state
$\ket{11 \ldots 11r}$, which has an error due to a remaining population
in the $\ket{r}$ state.
In the
example of Fig.~\ref{fig:setup}c), this fraction is below $10^{-5}$. 
When comparing the probability bars representing errors in the projective measurements,
the impact of crosstalk is evident as the bars differ by two orders of magnitude, i.e.,
up to $10^{-3}$ errors for $\rg = \sqrt{10} \al$ versus $10^{-5}$ for $\rg = \sqrt{16} \al$. 
The probabilities for this set of states shown in the histogram are extracted directly from the
TTN by sampling 1,000,000 projective measurements. For both $\rg=\sqrt{10} a$ 
and $\rg = \sqrt{16} a$, the samples cover at least $99.945\%$ of the probability,
i.e., a single measurement appears with at most $0.055\%$ probability not in the data shown;
these statistics can be directly compared to experiments.
Although simulations of
projective measurements are possible as shown in Fig.~\ref{fig:setup}c),
we concentrate in the remainder on the infidelity which is more accessible
in its interpretation as a single number.

\begin{figure}[t]
  \begin{center}
    \begin{minipage}{0.99\linewidth}
      \begin{overpic}[width=1 \columnwidth,unit=1mm]{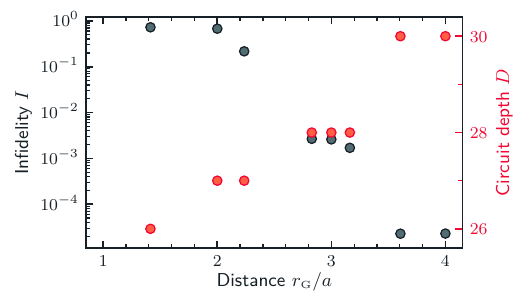}
        \put( 0,53){a)}
      \end{overpic}
    \end{minipage}
    \begin{minipage}{0.99\linewidth}
      \begin{overpic}[width=1 \columnwidth,unit=1mm]{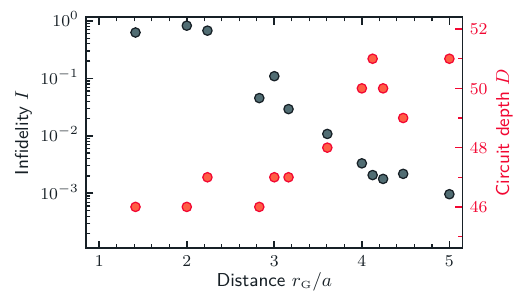}
        \put( 0,53){b)}
        \put(80, 0){\figshares{}}
      \end{overpic}
    \end{minipage}    \caption{\emph{Measuring the effect of controlled phase gates executed
        in parallel in a closed quantum system.}
      The infidelity $I$ decreases towards larger circuit depth $D$ for the GHZ
      state preparation.
      a)~For the $4 \times 4$ grid, we identify a clear jump for distances $r \ge 2$,
      which allows reducing the circuit depth by more than six percent allowing
      an infidelity of a $10^{-3}$ level.
      b)~For the $8 \times 8$ grid, larger distances have to be considered to
      go to a fidelity of the $10^{-2}$ level at $\rg \ge 4 \al$. A gain in circuit
      depth of $35\%$ is possible in comparison to a circuit without parallel
      CZ gates.
      \label{fig:effects_rydberg}}
  \end{center}
\end{figure}

Figure~\ref{fig:effects_rydberg} shows the change
in the circuit depth $D$ and the infidelity $I$ of the algorithm for preparing a global GHZ state as a function of the radius $\rg$.
The fidelity of the algorithm $F$ is defined as the state fidelity
$F = \left| \braket{\psi(\tau)}{\psi{\rms{GHZ}}} \right|^{2}$ at the end of
the algorithm at time $\tau$; the infidelity is $I = 1 - F$. 
The total time of the complete algorithm is $\tau = D \cdot 122 \mathrm{ns}$, which is the product of
the circuit depth $D$ times the $122 \mathrm{ns}$ pulse time per gate, and
the circuit depth $D$ depends itself on the compiler setting and system
size.
The circuit
is generated with the \RGHZC{} compiler targeting directly the
Rydberg platform. On the one hand,
the circuit depth drops, e.g., from over $51$ to $46$ while decreasing
the distance $\rg$ from $\sqrt{25} \al$ to $\sqrt{2} \al$ for the $8 \times 8$ square lattice.
On the other hand, we observe that the
infidelity changes by more than two orders of magnitude while
changing $\rg$. This change in infidelity is due to the
Rydberg interaction decaying with a power of six, especially
for the $8 \times 8$ grid where we have the bigger number of CZ gates $n$.
The overall fidelity $F$ depends on the number of CZ gates $n$, therefore
we define the average as $\mathcal{F}_{\rms \oslash} = F^{1/n}$. For the
largest values of $\rg$ shown in Fig.~\ref{fig:effects_rydberg}, the error is
driven by small numerical artifacts always remaining in an optimized pulse sequence, i.e.,
the value of $\mathcal{F}_{\rms \oslash}$ is in good agreement with the Bell state
fidelity for a single gate of the protocol for the closed system from Ref.~\cite{Pagano2022}.

We now choose the radius $\rg = \rs$ for a safely executed algorithm as
$\rs(L=4) = \sqrt{8} \al$ and $\rs(L=8) = \sqrt{16} \al$.
In Sec.~\ref{sec:rydberg}
on the Rydberg model, we show that the infidelity of the order of $10^{-3}$ introduced in
this scenario for a $4 \times 4$  grid is below the largest errors of the order of $10^{-2}$ introduced
by decay from the Rydberg state. The same argument holds for an infidelity of the $10^{-2}$
level for an $8 \times 8$ system.
In summary, Fig.~\ref{fig:effects_rydberg} demonstrates that we identify
the errors originating from parallel CZ gates on a Rydberg quantum computer;
the ideal setting for the system is $\rs(L=4) = \sqrt{8} \al$
and $\rs(L=8) = \sqrt{16} \al$, which leads to a tolerable loss in fidelity
in comparison to a circuit serial in the CZ gates. 
We point out the reduction of the circuit depth by $35\%$ for the
$8 \times 8$ system from a circuit depth 78 without any CZ gates in parallel to a circuit depth of 50, see Sec.~\ref{sec:rydberghz} for details. Meanwhile, the effect in the $4 \times 4$ system is almost
negligible and reduces the circuit depth from 30 to 28 or about 7\%, showing as expected a favorable scaling with the system size.
In Sec.~\ref{sec:rydberg}, we prove that reducing the circuit depth
and running gates in parallel allows one to minimize the overall error
from crosstalk and open system effects caused by imperfections in the trapping
of the atoms.

%

%
We now move on to the second example, which gives an outlook of parallel
circuits
beyond the NISQ applications: to the implementation of quantum
error-correcting codes~\cite{Gottesman2000,Wootton2018}. 
We construct a five-qubit repetition code encoding one logical qubit into five physical qubits
on a square lattice, which allows one to detect bit-flip errors.
Each logical qubit can be implemented with one physical qubit
and its four nearest neighbors. A repetition code needs to prepare the initial state
$\ket{\psi} = \alpha \ket{0} + \beta \ket{1}$ of each qubit in the algorithm into the
logical qubit $\ket{\Psi} = \alpha \ket{00000} + \beta \ket{11111}$. Without loss
of generality, we assume $\alpha = \beta = 1 / \sqrt{2}$ and a product state between
the logical qubits.
The preparation of the logical qubits can be parallelized to a greater extent
than the preparation of the global GHZ state on the complete lattice: the first
CZ gate of the global GHZ state has always to run in serial, i.e., there is only one
option for the control qubit of the first CZ gate; in contrast, this preparation of multiple local GHZ states
can have parallel gates from the first CZ gate on.
We have the option of a dense filling in the two-dimensional geometry or leaving
additional physical qubits unused, i.e., auxiliary
qubits in applications, between the logical qubits. We choose the first option
as the denser the packing is the more difficult the handling of crosstalk is.

\begin{figure}[t]
  \begin{center}
    \begin{minipage}{0.99\linewidth}
      \begin{overpic}[width=1.0 \columnwidth,unit=1mm]{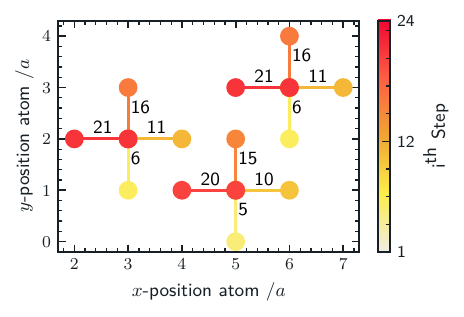}
        \put( 0,63){a)}
      \end{overpic}
    \end{minipage}\hfill
    \begin{minipage}{0.99\linewidth}
      \begin{overpic}[width=1.0 \columnwidth,unit=1mm]{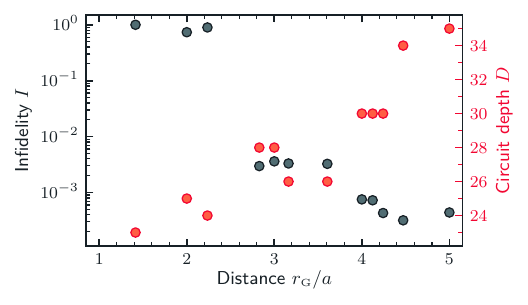}
        \put(0,53){b)}
        \put(82, -0){\figshares{}}
      \end{overpic}
    \end{minipage}
    \caption{\emph{Parallel initialization for quantum error-correcting codes.}
      a)~Parallel example algorithm for preparing three logical qubits of
      a five-qubit repetition code; links of the same color refer to CZ gates
      executed in parallel in the same layer. The underlying TTN has sixteen
      sites. The preparation requires 24 layers in total. The steps indicated
      by the color map depend on $\rg$, which is set to $\rg = 2 \al$ for this
      circuit.
      b)~The infidelity $I$ of GHZ state preparation for each logical qubit
      improves for larger radii $\rg$ and circuits depths $D$.
                                                                                \label{fig:qec}}
  \end{center}
\end{figure}

Figure~\ref{fig:qec}a) presents an example of the circuit for three
logical qubits in the dense filling scheme; three logical
qubits fit on a 16-site TTN. The binary TTN requires the number of
sites to be a power of two; additional sites inserted to reach the
next power of two do not affect the simulation, i.e., are in the
ground state and never addressed by a laser.
Each TTN site is still modeled as a qutrit.
The setup allows one to repeat
the analysis of the final fidelity as a function of $\rg$ in this scenario,
see Fig.~\ref{fig:qec}b). We want to ensure that one can reach the same level
of fidelity as in the global GHZ state, which is chosen with
regard to error sources from open quantum systems; ideally, we can use the same
distance $\rs$ than before to reach the same or a better infidelity. We
confirm the radius $\rs(L=4) = \sqrt{8} \al$
for a safe execution of the 15-qubit simulation of three logical qubits,
which reaches an infidelity of $10^{-3}$, which can be improved further by
one order of magnitude when increasing the radius further. Similarly, we
observe $\rs(L=8) = \sqrt{16} \al$ for the 60-qubit simulation with twelve
logical qubits reaching an infidelity of a $10^{-2}$ level.

\section{Modeling $^{88}$Sr quantum computers on two-dimensional grids          \label{sec:rydberg}}
%
The Rydberg quantum processor analyzed here assumes the following
characteristics: the atoms are trapped by optical tweezers in
a two-dimensional grid; it is possible to individually
address the neutral atoms, e.g., to drive the transition for the implementation of
single-qubit rotation gate $R_{\rms \scriptscriptstyle X}$ or the transition to the Rydberg state
for entangling two-qubit gates. The individual addressing of atoms allows one to
group several single-qubit gates and two-qubit gates together into groups, where each group of gates
can be executed at the same time. The strong interaction between the
atoms within the Rydberg blockade radius affects
the fidelity of executing controlled-phase gates in parallel.

We tailor the Hamiltonian towards an implementation with strontium-88
atoms, where the two-qubit states are encoded into the fine-structure qubit $\ket{0} = \ket{5^{3}P_{0}}$ and
$\ket{1} = \ket{5^{3}P_{2}}$; the Rydberg state $\ket{r} = \ket{60^{3}S_{1},m_{J} = 1}$ is required to implement the
controlled-phase gate. Overall, the Hamiltonian $\Hr$ and parameters follow
Ref.~\cite{Pagano2022} and consists of an idealized three-level system of
\begin{align}
  \Hr =& \sum_{j,k} \Omega_{j,k}^{x}(t) \sigma_{j,k}^{x}
               + \Omega_{j,k}^{z}(t) \sigma_{j,k}^{z} \nonumber \\
             & + \sum_{j,k} \left( \Omega_{j,k}^{R}(t) \ket{r} \bra{1} + h.c. \right) \nonumber \\
     & + \sum_{j,k} \sum_{j',k'} V(j,k,j'k') n_{j,k} n_{j',k'} \, ,
                                                                                \label{eq:ham}
\end{align}
where the qubits in the two-dimensional grid are indexed with $j$ and $k$. The
Hamiltonian contains both, an effective Hamiltonian for implementing single-qubit
gates and couplings to the interacting Rydberg states for the
implementation of the two-qubit CZ gates. The
effective Hamiltonian for the single-qubit gates uses the Pauli matrices
$\sigma^{x} = \ket{0} \bra{1} + \ket{1} \bra{0}$ for rotation-x gates of
an arbitrary angle and the $\sigma^{z} = \ket{0} \bra{0} - \ket{1} \bra{1}$
for rotation-z gates of an arbitrary angle. The single-qubit gates are driven by the
corresponding time-dependent effective pulses $\Omega^{x}$ and $\Omega^{z}$, respectively.
The single-qubit gates are executed with high fidelity with respect to other error
sources and are not the leading source of error~\cite{Shi2022}.
%
%
The transition
between the states $\ket{1}$ and $\ket{r}$ is driven by a single-photon
transition and represented in the Hamiltonian by the Rabi frequency
$\Omega^{R}$ and the transition $\ket{r} \bra{1}$ and
$\ket{1} \bra{r}$. The Rydberg interaction is modeled via
the interaction operator $n = \ket{r} \bra{r}$ and a van der Waals interaction,
where the strength depends on the distance $d$ as
\begin{align}                                                                           \label{eq:c6}
  V(j, k, j', k') =& \frac{-C_{6}}{d^6} \, ,
\end{align}
where $d = \sqrt{(x_{j} - x_{j'})^2 + (y_{k} - y_{k'})^2}$.
To allow for arbitrary lattice structures, the position of the qubit
labeled with the indices $(j, k)$ is $(x_{j}, y_{k})$.
The coefficient $C_{6}$ describes the strength of the van der Waals
interactions~\cite{Pagano2022}. 
This decaying interaction leads to the Rydberg blockade radius~\cite{Gaaetan2009,Urban2009}, in which
more than one excitation to the Rydberg state is prevented: the transition
from one excitation to two excitation is off-resonant due to an additional
energy shift. The Rydberg state is used for the implementation
of two-qubit gates and has an immediate effect
on the crosstalk of two gates running in parallel. The Hamiltonian of
Eq.~\eqref{eq:ham} is sufficient to implement a universal gate set consisting
of rotation-x, rotation-z, and CZ gates.

Finally, we include a Lindblad description of the system introducing
a decay of the Rydberg state $\ket{r}$ for the analysis provided
later in this section; as a starting point, we are interested
in how much our fidelity decreases under the assumption that all decays
end up in states outside our computational space,
i.e., the worst-case scenario. Therefore, we
include a non-Hermitian part to the Hamiltonian with the Lindblad operator
$L_{\rms{decay}} = \ket{d} \bra{r}$ analog to quantum
trajectories~\cite{Daley2014,QTNoJumps}, where $\ket{d}$ is any dark state
outside the computational states. We do not aim to retrieve $\ket{d}$ and
re-introduce it into the computational states. We obtain
for the open quantum system
\begin{align}                                                                   \label{eq:hoqs}
  \Ho &= \Hr - \mathrm{i} \gamma \sum_{j,k} \ket{r} \bra{r}_{j,k} \, .
\end{align}
From all the possible
error sources, the decay from the Rydberg state is the most important
source of error for a single CZ gate~\cite{Jandura2022,Pagano2022}. We
also choose the conservative estimate of
$1 / \gamma = 50 \mu\mathrm{s}$ for the decay time from Ref.~\cite{Pagano2022}.

In the following, we show that the remaining population in the Rydberg state
quantifies the crosstalk and can serve as an indicator of the fidelity of the
state preparation. The remaining population in the Rydberg state is
\begin{align}
  \PR &= \sum_{j,k} \langle n_{j,k}(t=\tau) \rangle \, ,
\end{align}
where we refer to the total time of the algorithm as $\tau$.
The integrated time in the Rydberg state $\TR$ is of interest when considering
the open quantum system with decay from the Rydberg state. Without crosstalk,
the time $\TR$ in the Rydberg state and final population in the Rydberg
state $\PR$ depend only on the number of CZ gates; we expect the same behavior
if the crosstalk is low enough. Therefore, $\TR$ and $\PR$ have a lower bound
which also cannot be improved by optimizations, where $\TR$ relates
to Eq.~\eqref{eq:hoqs}.
We define the integrated time in the Rydberg state as
\begin{align}
  \TR &= \sum_{j,k} \int_{0}^{\tau} \mathrm{d}t \langle n_{j,k}(t) \rangle \, ,
\end{align}
where the time $\TR$ connects directly to the probability of loss from
the Rydberg state to a state outside of the computational space; this
fraction of states identified by the lost norm is accounted for with
a fidelity of zero.

\begin{figure*}[t]
  \begin{center}
    \begin{minipage}{0.49\linewidth}
      \begin{overpic}[width=1.0 \columnwidth,unit=1mm]{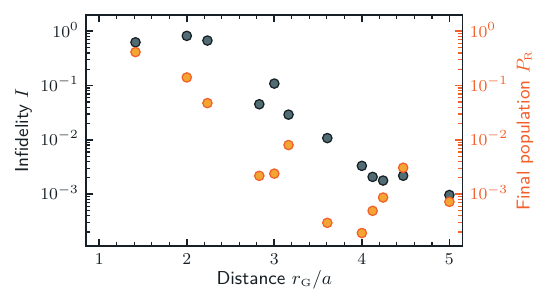}
        \put( 0,52){a)}
      \end{overpic}
    \end{minipage}\hfill
    \begin{minipage}{0.49\linewidth}
      \begin{overpic}[width=1.0 \columnwidth,unit=1mm]{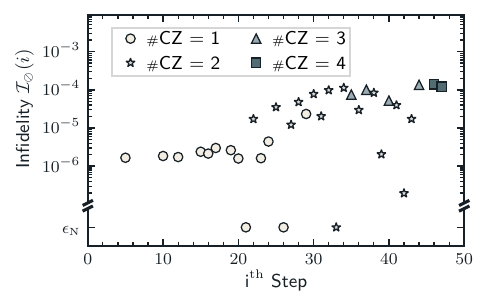}
        \put( 0,58){b)}
      \end{overpic}
    \end{minipage}
    \begin{minipage}{0.49\linewidth}
      \begin{overpic}[width=1.0 \columnwidth,unit=1mm]{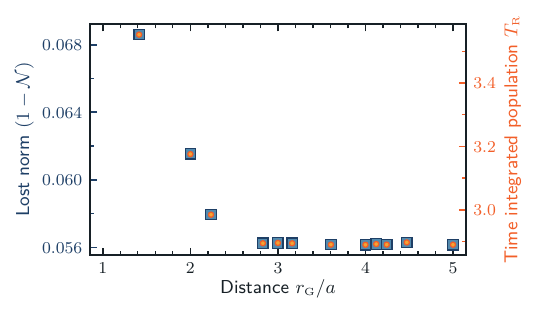}
        \put( 0,53){c)}
      \end{overpic}
    \end{minipage}\hfill
    \begin{minipage}{0.49\linewidth}
      \begin{overpic}[width=1.0 \columnwidth,unit=1mm]{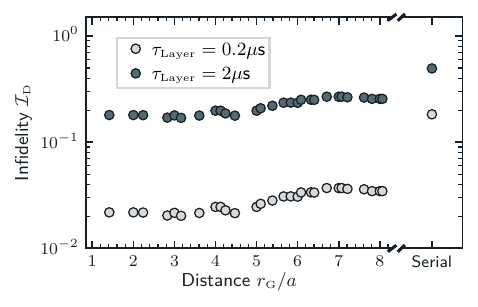}
        \put( 0,58){d)}
        \put(77, 0){\figshares{}}
      \end{overpic}
    \end{minipage}
    \caption{\emph{Rydberg measurements for GHZ state preparation executed in parallel on $8 \times 8$ systems.}
      a)~We consider the populations in the Rydberg state for the final
      state of a closed system. Running parallel gates
      close to each other results in a large population in the Rydberg state,
      which causes an increase of the infidelity.
      b)~We measure the infidelity after the application of each parallel
      layer and calculate the average error per CZ in one layer,
      i.e., $\mathcal{I}_{\rms{\oslash}}(i) = \mathcal{I}(i)^{-n}$ for
      $n$ CZ in the i$^{\mathrm{th}}$ layer. Three data points cannot
      be captured by the numerical precision $\epsilon_{\rms{N}}$ of the
      simulation and appear on the $y$-axis as $\epsilon_{\rms{N}}$.
      c)~The open quantum system causes a finite loss of norm and
      time-integrated Rydberg population $\TR$ because the execution
      of CZ gates temporarily populates the Rydberg state. But $\TR$ becomes
      even larger at small radii $\rg$ due to permanent Rydberg populations
      related to running parallel CZ gates in proximity to each other.
      d)~The estimate for the infidelity introduced due to
      dephasing shows a jump of one order of magnitude for the serial circuit
      versus the parallel circuit as well as for the comparison between the two
      scenarios with different layer durations $\tlayer$.
                                                                    \label{fig:meas_rydberg}}
  \end{center}
\end{figure*}

With these observables, we analyze the choice of the minimal radius of $\rg$
and to which extent we trade a better fidelity for a larger circuit depth.
The minimal radius $\rg$ describes the minimal distance between any two qubits
of different CZ gates executed in parallel, e.g., a distance of $\rg = 2 \al$
in Fig.~\ref{fig:setup}a).
We recall the dependency of the fidelity and the minimal radius $\rg$ for
entangling gates running in parallel discussed in
Fig.~\ref{fig:effects_rydberg}, and we move towards looking at
measurement specific to the Rydberg state of our platform in the next paragraph.

We observe that the parallel execution of CZ gates leads to a remaining
population in the Rydberg state $\PR$ as the gate is designed for serial use,
i.e., the laser pulse used to drive the gate has been engineered for two perfectly isolated Rydberg atoms.  In contrast, the system encounters four additional two-body interactions with interaction
strengths depending on the positions of the atoms when running two CZ gates close to each other; then, the pulse sequence fails to perfectly reproduce
the CZ gates. The remaining population in the Rydberg state $\PR$
presents a path to measure the
effects of crosstalk and is shown in Fig.~\ref{fig:meas_rydberg}a).
We employ the fact that 
$\ket{r}$ is solely used for the gate
implementation. Therefore, the measurement of the population in the
Rydberg state is experimentally possible: after a readout of the
qubit states $\ket{0}$ and $\ket{1}$ the probability remains below
$1$, which is one way to estimate the population in the Rydberg
state $\ket{r}$ and related losses. This two-state readout scheme can be implemented
via an additional state, see Fig.~\ref{fig:setup}c)i).

The digital twin allows one to evaluate which layers contribute
the most to the final infidelity. Figure~\ref{fig:meas_rydberg}b)
refines the average infidelity per CZ gate on the level of the
i$^{\mathrm{th}}$ parallel layer for a closed quantum system, i.e., $\mathcal{I}_{\rms \oslash}(i)$.
The underlying average gate fidelity of the CZ gate is optimized up to
$\mathcal{F}_{\rms{CZ}} = 99.99983\%$ and represents a meaningful reference
value for $1 - \mathcal{I}_{\rms \oslash}(i)$.
Although the
error depends on the distances of parallel gates, we interpret
the trend to higher errors at the end of the circuit as a sign
that errors propagate, e.g., a small population in the Rydberg
state from a previous layer for the control qubit of the CZ gate
affects the next layer. A difference between the system sizes
is the number of previous errors that can affect CZ gates at a later
stage, i.e, the paths for executing consecutive CZ
gates in the $8 \times 8$ grid are longer than in the $4 \times 4$
grid, which is another reason for smaller average fidelities
$\mathcal{F}_{\rms \oslash}$ for the $8 \times 8$ grid, in addition
to running a greater fraction of CZ gates in parallel.

The accumulated infidelity $I$ due to the crosstalk of the parallel
algorithm has to be compared to other experimentally relevant
sources of error. The decay from the Rydberg state has been
identified as the major source of error for single-qubit and two-qubit gates~\cite{Jandura2022,Pagano2022}.
We assume the worst-case scenario that any decay from the Rydberg
state $\ket{r}$ leaves our computational basis of the states
$\ket{0}$, $\ket{1}$, and $\ket{r}$. The non-Hermitian version
of the Hamiltonian $\Ho$ in Eq.~\eqref{eq:hoqs}
describes this scenario of an open quantum
system; the remaining norm $\mathcal{N}$ of the state can be directly
multiplied with the final fidelity to obtain the average
fidelity of any quantum algorithm. For the square lattice
with 16 qubits in total, we obtain a loss of norm in the order
of $1.4 \cdot 10^{-2}$, i.e., a final fidelity of about $98.6\%$ in the limit of large $\rg$
assuming an otherwise perfect circuit. The lost norm is a good approximation
of the infidelity, because the decay is the leading source of error for
large radii $\rg$. Therefore, the error
originating from crosstalk with $\rg(L=4) = \sqrt{8} \al$ below the order of
$3 \cdot 10^{-3}$ or less is negligible. We show the data for 64 qubits
in Fig.~\ref{fig:meas_rydberg}c), where the fidelity drops to about
$94.4\%$, i.e., an infidelity on the $5.6 \cdot 10^{-2}$ level.  We point out that increasing
the radius $\rg(L=8)$ beyond $\sqrt{9} \al$ in the regime of
high fidelities leaves the loss in norm unchanged. Recall that the
Lindblad operator acts on the state $\ket{r}$, which is
the only state contributing
to this loss in norm and we can demonstrate this connection by measuring
$\TR$, i.e., the integrated population in the state $\ket{r}$ for
all qubits during the complete evolution, as shown in the right
y-axis of Fig.~\ref{fig:meas_rydberg}c). For the same system size,
we reach infidelities below $5 \cdot 10^{-3}$ at $\rg = \sqrt{16} \al$
and therefore we set $\rs = \sqrt{16} \al$ as the error from the decay
is dominating over the error from the crosstalk.

The challenge of conserving coherence in the quantum system is not
reflected in the non-Hermitian Hamiltonian of Eq.~\eqref{eq:hoqs},
i.e., the open system term introduced with the decay does not favor
circuits with shorter total time as one expects for the actual
hardware.
After the leading contribution of decay from the Rydberg state, the
next relevant term stems from the fluctuations around the magic
trapping condition~\cite{Li2019a,Meinert2021}, which lead to decoherence, especially in the
case of the GHZ state. 
In principle, perfect magic trapping is designed so that all three levels pick up the same phase and dephasing is eliminated. However, the magic trapping condition is never met perfectly in an experimental realization, e.g., because of field noise.
We include this effect aposteriori into the
results and consider how the decoherence time $T_{2} = 10 \mathrm{ms}$
affects the fidelity over the time $t$.
The fidelity between
a GHZ state of $n$ qubits exposed to dephasing and
a perfect GHZ state is
\begin{align} \label{eq:dephasing}
  \Fdephasing{}(t) = \frac{1}{2} + \frac{1}{2} \exp\left( - \frac{n \cdot t}{T_{2}} \right) \, .
\end{align}
To include this effect in our estimate, we approximate that
this decoherence affects each qubit starting with their first
$\RotG{X}{}{\phi}$ gate; to simplify the estimate, we neglect
an additional treatment of the $\ket{r}$ state as it appears
only in a limited fraction of steps while applying the CZ gates.
The pulse time for a
CZ is $0.122\,\mu\text{s}$ and we choose the amplitudes for
the single-qubit gates to match the time of the CZ gate.
Recall that the qubit
is encoded into the fine-structure qubit and thus single-qubit
gate within $0.122\,\mu\text{s}$ are possible because of the
strong Raman transition. However, this duration of the pulses
neglects intermediate steps required between layers. Therefore, we follow two
scenarios: the first scenario runs one layer every $0.2\,\mu\text{s}$
allowing for a gap after each pulse, and the more conservative second
scenario with one layer every $2\,\mu\text{s}$, e.g., including the possibility
to move the addressing laser beams to different atoms between layers.
Figure~\ref{fig:meas_rydberg}d) proves the need to parallelize
the circuit. For both serial circuits and the parallel circuits
with $\tlayer = 2\,\mu\text{s}$, the decoherence is the leading
source of error, i.e., even more important than decay from the Rydberg
state. All of the circuits with $\tlayer = 0.2\,\mu\text{s}$
with parallel local gates bring the decoherence error below the
level of the decay from the Rydberg state. The parallel circuits
fluctuate due to the freedom of the scheduler, but the data
suggests a trend towards higher errors from decoherence for
larger circuit depths $D$. There are no possible distance
in an $8 \times 8$ system beyond $\rg / a > \sqrt{74} = \sqrt{7^2 + 5^2} \approx 8.6$ which still have
parallel gates; but in general, the serial case is with $\rg / a = \infty$.

\begin{figure}[t]
  \begin{center}
      \begin{overpic}[width=1.0 \columnwidth,unit=1mm]{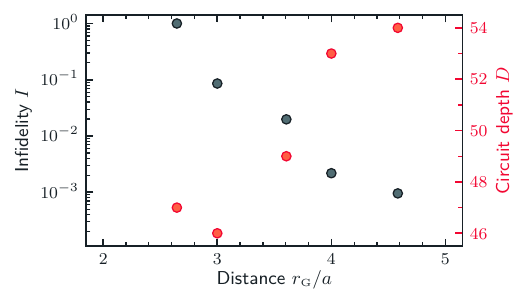}
        \put(80, 0){\figshare{}}
      \end{overpic}
    \caption{\emph{Hexagonal lattices with $8 \times 8$ Rydberg atoms.}
      Infidelities $I$ for the GHZ state preparation on a hexagonal
      lattice of a closed quantum system.
                                                                                \label{fig:hex_rydberg}}
  \end{center}
\end{figure}

Rydberg atoms are able to implement other lattice layouts than
the square lattice, e.g., a hexagonal lattice. Although a hexagonal
lattice has advantages in terms of connectivity, the parallelization of
gates is actually more subtle due to the denser packing of the atoms
in the lattice. For example, the area of a rectangle containing the Rydberg
atoms of an $8 \times 8$ hexagonal lattice is ten percent smaller than
the area of the square containing $8 \times 8$ atoms of a square lattice.
The denser packing leads to the fact that more atoms are within the
Rydberg blockade radius and parallelization is more restricted.
Figure~\ref{fig:hex_rydberg} provides an impression of the infidelity
as a function of $\rg$ in the hexagonal lattice layout. Overall, we reach
similar fidelities than in the square lattice; the circuit depths $D$ tend
to be higher, while the $\rg$ has the same effect on the infidelity. For
example, a $10^{-2}$ fidelity is reached in both layouts for $3 \al < \rg < 4 \al$.

Altogether, the combination of the parallel GHZ state preparation with
the characteristics of the specific Rydberg setup allow us to obtain
detailed insights, e.g., experimentally feasible measurements to characterize
the error of crosstalk or open system effects, and explore the consequences
of different lattice layouts. Although the infidelity estimated for the
decoherence fluctuates due to different scheduling, a comparison to the
serial circuit shows the necessity to run algorithms in parallel and minimize
the circuit depth as long as the crosstalk is controllable.


\section{Tensor network simulations}
                                                                                \label{sec:algo}
%
The  digital twin is based on the tensor network
simulations: here, we 
employ a binary TTN based on the software framework \emph{Quantum TEA}~\cite{Silvi2017,Felser2021,Ballarin2021,QuantumTEAPackage}
and account for the two-dimensional architecture of the Rydberg array
via the Hilbert curve mapping: the Hilbert curve maps a square lattice
into a one-dimensional system and is in many scenarios favorable in
comparison to other schemes~\cite{Cataldi2021}. Although the TTN is
better suited for the two-dimensional system than a matrix product state,
we still require a mapping and an ordering with a single index to label the
leaves of the tree network. We display the original
$4 \times 4$ lattice before its mapping and the tree network in Fig.~\ref{fig:setup}a).
The time evolution is executed
via the time-dependent variational principle~\cite{Haegeman2016,Bauernfeind2020} and a two-tensor update,
supporting long-range interactions required for the Rydberg
interactions in Eq.~\eqref{eq:ham}. Due to the underlying two-dimensional
grid and mapping with the Hilbert
curvature, the Hamiltonian is described via a tensor product
operator, i.e., a set of two-body interactions~\cite{Felser2021PhD}, and
the number of two-body interaction terms grows as we increase
the radius of Rydberg interactions that should be included. In the limit
of an all-to-all interaction, the number of terms in the tensor product operator representing the Hamiltonian
scales quadratically with the number of atoms. The translation of
the gate sequence into the time evolution is split into each parallel layer
to simplify the control of the time-dependent parameters. For example,
the time-dependent laser-driving from Eq.~\eqref{eq:ham} are set to zero
for atoms not involved in any gate in the specific layer; atoms involved
in a gate have exactly one out of $\Omega^{x}$, $\Omega^{z}$, and
$\Omega^{R}$ turned on for their layer as each of them drives a different
gate. The local gates are implemented as square pulses, the $CZ(\phi)$
gate as described in Ref.~\cite{Pagano2022}.

Tensor network simulation presents a classical path toward many-body quantum
simulations due to an entanglement-driven approximation: the simulation can become
exact at a certain bond dimension $\chi$ controlling the maximal amount
of bipartite entanglement given by the Hilbert space or if the relevant quantum state does not exceed the entanglement
that can be described with $\chi$. The GHZ state of our algorithm represents
such a state with a low bond dimension, i.e., the state can be written with
$\chi = 2$. Although we pick up additional entanglement due to imperfections
of the gates, the target state represents the ideal playground to exploit
tensor network in this scenario of providing results of a quantum computer
up to the pulse level. We recall that even the $4 \times 4$ grid of the
three-level system corresponds to an equivalent of simulating about
$25$ qubits.
The square lattice of $8 \times 8$ qutrits is equivalent to about
$101$ qubits and beyond the reach of exact methods and their upper
limit of around 45 qubits~\cite{JaschkeED,Haner2017}.

The long-range interactions of the Rydberg atoms play a crucial role.
Therefore, we include more than the nearest-neighbor interactions at distances
$\al$, which are required for the CZ gate in $x$- and $y$-direction. We incorporate
additional interactions depending on the radius $\rg$ for running two CZ gates
in parallel: interactions up to $\ri = \rg + \doff$ are included. Therefore, the
computational effort increases for large $\rg$ with respect to the number of
terms in the Hamiltonian and the total number of time steps. The latter
is induced  by larger circuit depths. Due to the
possibility of the remaining population in the Rydberg state $\ket{r}$, we are
required to keep all interactions up to the radius $\ri$. This required
range of interaction in the Hamiltonian is a direct consequence of Eq.~\eqref{eq:c6}
and the Rydberg blockade radius.

In summary, there are three crucial parameters for the convergence study
presented in Tab.~\ref{tab:conv8x8}, i.e., the bond dimension $\chi$, the
time step $dt$ of the evolution, and the range of interactions $\ri$ included
in the Hamiltonian, where $\ri = \rg + \doff$ is tuned via $\doff$. To prove
convergence, we validate the infidelity of the final states obtained by two
different parameters $p_1$ and $p_2$ for either $\chi$, $dt$, or $\doff$ and their
effect on the fidelity with the perfect GHZ state, i.e.,
\begin{align}
    \mathcal{C}(p_{1}, p_{2}) &= 1 - \left|\braket{\psi(\tau, p_{1})}{\psi(\tau, p_{2}} \right|^{2} \, , \\
    \Delta_{\mathcal{F}}(p_{1}, p_{2}) &= \mathcal{F}(p_{1}) - \mathcal{F}(p_{2}) \, .
\end{align}
The intermediate
regime with a limited amount of crosstalk is the most meaningful for the convergence
study. In the limit of small $\rg$, the compiler produces as expected circuits with low fidelity and a convergence study in this regime does not require long-range interactions to the same extent as in the parameter regime suggested by us as $\rs$ in Sec.~\ref{sec:parallel}. The
circuit in the limit of large $\rg$ has no crosstalk and matches the GHZ state
up to the accumulated errors of a CZ; this scenario does not challenge the convergence
with regard to the bond dimension nor the radius $\ri$. Therefore, we check
convergence for the $8 \times 8$ grid for $\rg = \sqrt{10} \al$ as well as
$\rg = \rs = \sqrt{16} \al$, where both $\mathcal{C}(p_{1}, p_{2})$ and
$\Delta_{\mathcal{F}}(p_{1}, p_{2})$ stay below a $10^{-3}$ level for
increasing the bond dimension $\chi$ by fifty percent, decreasing the time
step by a factor of $10$, and increasing $\ri$ by four lattice spacings $a$, except
for one data point for $\rg = \sqrt{10} \al$. One additional intermediate data point
is shown for comparison. If we compare $\mathcal{C}(p_{1}, p_{2})$ and
$\Delta_{\mathcal{F}}(p_{1}, p_{2})$ to the infidelities
of Fig.~\ref{fig:effects_rydberg}b), the numerical error stays about one
order of magnitude below the corresponding infidelities for $\rg = \sqrt{10} \al$
and $\rg = \sqrt{16} \al$, respectively. Hence, we consider the results converged
on the relevant order of magnitude and use $\chi = 12$, $dt = 0.001 \mu\mathrm{s}$,
and $\doff = 2 \al$ for all simulations.

\begin{table}[t]
  \begin{center}
  \caption{\emph{Convergence study for tensor networks.} We evaluate the
    convergence of the tensor network simulations as a function of
    the bond dimension $\chi$, the time step $dt$, and the radius $\ri$
    of the interactions included in the Hamiltonian. We focus on the
    overlap of the final states of simulations with respect to the
    parameters $p_{i}$ that we check in convergence, i.e.,
    $\mathcal{C}(p_{1}, p_{2}) = 1 - \left|\braket{\psi(\tau, p_{1})}{\psi(\tau, p_{2}} \right|^{2}$,
    and the induced change in the fidelity to the GHZ state, i.e.,
    $\Delta_{\mathcal{F}}(p_{1}, p_{2}) = \mathcal{F}(p_{1}) - \mathcal{F}(p_{2})$.
    Distances are implicitly given in terms of the lattice spacing
    $a$ and the time in $\mu\mathrm{s}$. Recall that the perfect GHZ
    state serving as the target has a bond dimension of $\chi = 2$.
                                                                                \label{tab:conv8x8}}
  \begin{tabular}{@{} lcc @{}}
    \toprule
    Convergence $\chi$
    & $\rg = \sqrt{10} \al$
    & $\rg = \sqrt{16} \al$ \\
    \cmidrule(r){1-1} \cmidrule(rl){2-2} \cmidrule(l){3-3}
    $\;\; \mathcal{C}(\chi=12, \chi=18)$ & $0.43 \cdot 10^{-3}$ & $0.13 \cdot 10^{-3}$ \\
    $\;\; \mathcal{C}(\chi=15, \chi=18)$ & $0.48 \cdot 10^{-3}$  & $0.10 \cdot 10^{-3}$ \\
    $\;\; \Delta_{\mathcal{F}}(\chi=12, \chi=18)$ & $-0.05 \cdot 10^{-3}$ & $-0.01 \cdot 10^{-3}$ \\
    $\;\; \Delta_{\mathcal{F}}(\chi=15, \chi=18)$ & $0.26 \cdot 10^{-3}$ & $-0.01 \cdot 10^{-3}$ \\
    \cmidrule{1-3}
    Convergence $dt$
    & $\rg = \sqrt{10} \al$
    & $\rg = \sqrt{16} \al$ \\
    \cmidrule(r){1-1} \cmidrule(rl){2-2} \cmidrule(l){3-3}
    $\;\; \mathcal{C}(dt=0.001, dt=0.0001)$ & $0.78 \cdot 10^{-3}$ & $0.24 \cdot 10^{-3}$ \\
    $\;\; \mathcal{C}(dt=0.005, dt=0.0001)$ & $0.80 \cdot 10^{-3}$  & $0.23 \cdot 10^{-3}$ \\
    $\;\; \Delta_{\mathcal{F}}(dt=0.001, dt=0.0001)$ & $0.20 \cdot 10^{-3}$ & $0.12 \cdot 10^{-3}$ \\
    $\;\; \Delta_{\mathcal{F}}(dt=0.005, dt=0.0001)$ & $0.31 \cdot 10^{-3}$& $0.08 \cdot 10^{-3}$ \\
    \cmidrule{1-3}
    Convergence $\ri = \rg + \doff$
    & $\rg = \sqrt{10} \al$
    & $\rg = \sqrt{16} \al$ \\
    \cmidrule(r){1-1} \cmidrule(rl){2-2} \cmidrule(l){3-3}
    $\;\; \mathcal{C}(\doff = 2, \doff = 6)$  & $0.36 \cdot 10^{-3}$ & $0.13 \cdot 10^{-3}$ \\
    $\;\; \mathcal{C}(\doff = 4, \doff = 6)$  & $0.54 \cdot 10^{-3}$ & $0.13 \cdot 10^{-3}$ \\
    $\;\; \Delta_{\mathcal{F}}(\doff = 2, \doff = 6)$ & $1.5 \cdot 10^{-3}$  & $0.25 \cdot 10^{-3}$ \\
    $\;\; \Delta_{\mathcal{F}}(\doff = 4, \doff = 6)$ & $0.32 \cdot 10^{-3}$  & $0.02 \cdot 10^{-3}$ \\
    \bottomrule
  \end{tabular}
  \end{center}
\end{table}

\section{Compiler for global GHZ states}                              \label{sec:rydberghz}
%
The compiler \emph{\RGHZC{}} 
compiles the algorithm for the global GHZ state preparation for the
given constraints: the connectivity preventing gates within the
radius $\rg$, the available gate set consisting of single-qubit rotations
around the $x$-axis and the $z$-axis, and the CZ gate. We choose to
implement a compiler from scratch due to the very specific target state
and the parameters we need to tune, i.e., the distance $\rg$ for
parallel gates. Our compiler has two modes: the first mode outputs
an algorithm with one initial Hadamard gate and a sequence of CNOT gates;
the second mode targets directly a subset of native gates
available on the digital twin of the Rydberg platform.
We focus on the latter in the following, because the first mode is a special
case of the second mode which is explained after the details of the algorithm.

The CNOT gate on a
Rydberg platform can be executed as $H_2 CZ_{1,2} H_2$ where the
Hadamard gate $H$ acts on the target atom~\cite{Henriet2020}. We instead use
the following decompositions for the Hadamard and $\mathrm{CZ}(\phi)$ gate obtained
by a derivate of the open-source qoqo compiler adapted for
the Rydberg platform~\cite{qoqo}
\begin{align}
  H =& \RotG{Z}{}{\frac{\pi}{2}} \RotG{X}{}{\frac{\pi}{2}} \RotG{Z}{}{\frac{\pi}{2}} \, ,
\end{align}
\begin{align}
  \mathrm{CNOT} =& \RotG{X}{C}{\pi}
        \RotG{X}{T}{\frac{\pi}{2}}
        \RotG{Z}{T}{-\frac{\pi}{2}}
        \RotG{X}{T}{\frac{\pi}{2}}
        CZ(\phi) \nonumber \\
      & \RotG{Z}{C}{\phi - \frac{3 \pi}{2}}
        \RotG{X}{C}{\pi}
        \RotG{Z}{C}{\frac{3\pi}{2}} \nonumber \\
      & \RotG{Z}{T}{- \phi - \frac{3 \pi}{2}}
        \RotG{X}{T}{\frac{\pi}{2}}
        \RotG{Z}{T}{\frac{3 \pi}{2}}
        \, .
\end{align}
In practice, the experimental implementation of a $\mathrm{CZ}(\phi)$
gate represents the CZ gate up to single-qubit rotations~\cite{Levine2019}.
The controlled-phase gate $\mathrm{CZ}(\phi)$ has therefore more gates
in comparison to the previously mentioned decomposition of Ref.~\cite{Henriet2020}.
We distinguish the control qubit and the target qubit by a superscript
\guillemotleft{}C\guillemotright{} and \guillemotleft{}T\guillemotright{}, respectively.

\begin{figure}[t]
  \begin{minipage}{0.37\linewidth}\raggedright a)
  \end{minipage}\hfill
  \begin{minipage}{0.60\linewidth}\raggedright b)
  \end{minipage}\hfill
  \begin{center}
    \begin{minipage}{0.37\linewidth}
      \begin{overpic}[width=1.0 \columnwidth,unit=1mm]{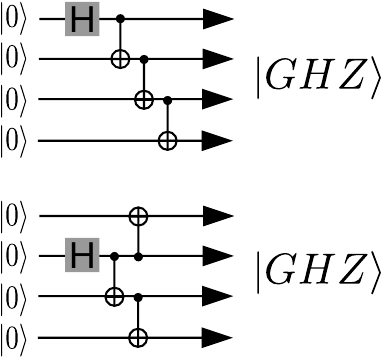}
      \end{overpic}
    \end{minipage}\hfill
    \begin{minipage}{0.60\linewidth}
      \begin{overpic}[width=1.0 \columnwidth,unit=1mm]{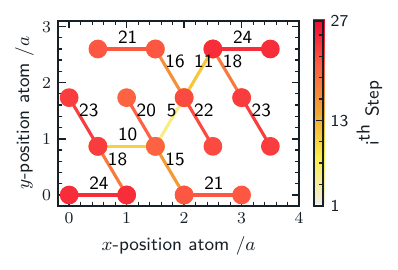}
      \end{overpic}
    \end{minipage}\hfill
   \end{center}
   \begin{center}
     \begin{minipage}{0.98\linewidth}
      \begin{overpic}[width=1.0 \columnwidth,unit=1mm]{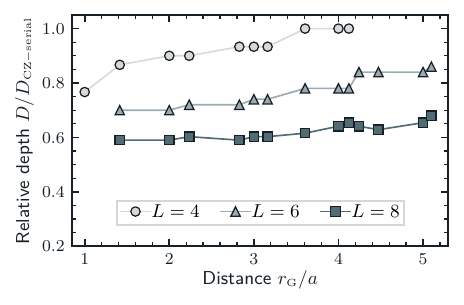} 
        \put( -1,61){c)}
        \put(80, 0){\figshares{}}
      \end{overpic}
    \end{minipage}
    \caption{\emph{Circuit depth of a GHZ state preparation.} a)~The upper
      circuit shows the serial preparation of the GHZ state in a one-dimensional
      geometry. The lower half sketches the parallel approach for
      the preparation of a one-dimensional system with nearest-neighbor
      connectivity and the algorithm follows a light cone starting in the middle and moving
      outwards.
      b)~Example for a parallel
      preparation of the GHZ state in a $4 \times 4$ hexagonal grid where the minimal distance
      between sites of two CZ gates is set to $\rg = 2 \al$. Links of the same color are CZ gates
      executed in parallel; the total circuit depth for the Rydberg platform is $D = 27$.
      The positions of the atoms are shown in terms of the lattice spacing $a$.
      c)~Selecting a minimal distance $\rg$ between the sites of two
      CZ gates to avoid the crosstalk of the Rydberg blockade leads to
      different circuit depths; truncations of intermediate
      states are set for all system sizes. We point out that the speed-up
      from a parallel execution grows with increasing system sizes.
                                                                                \label{fig:circuitdepth_GHZ}}
  \end{center}
\end{figure}

Our approach to compiling the algorithm relies on the fact that all
qubits are initially in the state $\ket{0}$; moreover, any qubit already
part of the GHZ state can act as a control qubit in an entangling gate to
add qubits to the GHZ state multiple times.
Figure~\ref{fig:circuitdepth_GHZ}a) summarizes the idea of how to
parallelize the algorithm in a one-dimensional system reducing the
circuit depth $D$ to $(L / 2 + 1)$ for even $L$; the benefit can be even
larger in a two-dimensional square system where the circuit depth of $L^{2}$
in the serial execution can be reduced to $(L + 1)$ for even $L$. In both
examples, we consider the Hadamard gate combined with CNOTs and no
restriction on the number of parallel gates executed at the same time
with a nearest-neighbor connectivity. The Rydberg platform has the lower
bound of $\Dmin = (3 + 5 L)$ for even $L$ and $\rg = \al$; due to $\rg = \al$, this
remains a theoretical lower bound without an acceptable fidelity~\cite{DminBound}.
%
%
The meaningful upper bound is an algorithm without any parallel CZ gates,
while parallel local gates are allowed: $\Dczser = L^2 + 14$. This upper bound
is $30$ and $78$ for a $4 \times 4$ and $8 \times 8$ grid, respectively.
The completely serial execution of all gates is $\Dser = 11 (L^2 - 1) + 3$.

%

Our compiler lists all the possible intermediate configurations of a GHZ circuit
up to a $4 \times 4$ grid in a reasonable time.
The number of intermediate configurations rapidly grows with increasing
system sizes, which forces us to truncate the least promising 
configurations based on metrics defined in the following.
The compiler without truncation contains the following steps:
\begin{enumerate}[noitemsep,nolistsep]
  \item{Fix minimal distance $\rg$, i.e., the minimal
    distance between any two atoms participating in different CZ
    gates at the same time.}
  \item{A set of unique sites is selected as initial starting points for the
     Hadamard gate and store them as a list of configurations
     $\mathcal{C}_{\gamma}$. No unique site inside the set maps via
     rotations or reflections of the grid to another site within the set.}
  \item{List all possible pairs $\mathcal{P}_{\alpha}$ of control and target sites for the
    configurations $\mathcal{C}_{\gamma}$.}
  \item{List all sets $\mathcal{S}_{\beta}$ of pairs $\mathcal{P}_{\alpha}$ which
    can run in parallel while obeying the minimal distance $\rg$.
   Subsets are not included, e.g., if 
   $\mathcal{S}_{\beta} = (\mathcal{P}_{A}, \mathcal{P}_{B}, \mathcal{P}_{D})$
   is included, we omit $\mathcal{S}_{\beta'} = (\mathcal{P}_{A}, \mathcal{P}_{B})$.}
  \item{Create a new configuration $\mathcal{C}_{\gamma}'$ for each $\mathcal{S}_{\beta}$. Check
    if the new configuration $\mathcal{C}_{\gamma}'$ is a solution.}
  \item{Block the qubits involved in a CZ gate for the next four iterations
    to allow the application of the local gates for the Rydberg platform.
    Reduce iterations blocked for local operations by one for all qubits
    not involved in a CZ gate.}
  \item{Continue with step 3) using the new configurations
    $\mathcal{C}_{\gamma}'$ as $\mathcal{C}_{\gamma}$. Allow to truncate
    entries in $\mathcal{C}_{\gamma}$ based on symmetry arguments
    or truncation rules.}
\end{enumerate}
The scaling of the number of intermediate configurations is kept to a minimum
due to the exploitation of symmetries, e.g., step 2) generates only three
unique configurations for a $4 \times 4$ square lattice. The other configurations are
rotations and reflections of those three configurations. Despite this effort
throughout the algorithm, we require the truncation of configurations for larger system
sizes.

The most convenient place for the truncation of configurations is
step 7) before starting the next iteration. We assign each configuration
to a bin according to the number of already entangled qubits and use it as the first criterion:
we keep truncating the bin with the smallest number of qubits and the configurations
therein as long as we keep a given percentage of configurations defined
as a parameter. The second metric takes into account the geometry of the
intermediate configuration. We consider a center-of-mass close to the
center of the grid as favorable because all corners can be reached in an
equal number of steps. Moreover, a configuration with a large average
distance from its own center of mass is closer to the corners of the system
and comes with a higher chance to run gates in parallel in the next
step. The fraction of configurations resulting in the largest intermediate
GHZ state is at least kept with respect to the previous configuration.
An option for a more aggressive truncation keeps only the largest
sets $\mathcal{S}_{\beta}$ in step 4), which stays disabled for the
Rydberg system.

Considering the compiler for the subset of native Rydberg gates, we point out
how the first compiler mode in terms of a Hadamard gate and CNOT gates is working:
consecutive CNOT gates can run without intermediate single-qubit gates,
which results in setting the blocked cycles in step 6) to zero for \RGHZC{}.

\begin{table}[t]
  \begin{center}
  \caption{\emph{Statistics for GHZ circuit.} Given a safe distance $\rs = \rg$
     to run CZ gates in parallel, we evaluate the circuit depth and number of
     single-qubit and two-qubit gates. In detail, we show the lower bound
     of the circuit depth $D_{\min}$ at $\rg = \al$ for a nearest-neighbor
     connectivity, the actual circuit depth $D(\rs)$, the circuit depth
     with all CZ gates in serial $\Dczser$, and the circuit depth
     with all gates in serial $\Dser$, i.e., the number of gates.
     Furthermore, we provide the average number $O_{1}$ and $O_{2}$ as 
     well as the maximal number MAX$_{1}$ and MAX$_{2}$ per layer of
     one-qubit and two-qubit gates, respectively. These numbers for the choice of $\rs$
     give an estimate of the hardware requirements, e.g., in terms
     of laser power. Finally, the Quantum Gates per Second (QGS) enable
     a comparison across different platforms, where we present two
     scenarios with $\tlayer = 200 \mathrm{ns}$ per layer and
     $\tlayer = 2 \mu\mathrm{s}$.
                                                                                \label{tab:statistics}}
  \begin{tabular}{@{} cccc @{}}
    \toprule
    Circuit property        &    $\;4 \times 4 \; (16)\;$ &     $\;6 \times 6 \; (36)\;$ &     $\;8 \times 8 \; (64)\;$      \\
    \cmidrule(r){1-1}      \cmidrule(rl){2-2} \cmidrule(rl){3-3} \cmidrule(l){4-4}
    $\rs$ &                $\sqrt{8} \al$ &       $4 \al$ &                $4 \al$                 \\
    $\Dmin$ &              23 &               33 &               43                \\
    $\Ds$ &                28 &               39 &               50                \\
    $\Dczser$ &            30 &               50 &               78                \\
    $\Dser$ &              168 &              388 &              696               \\
    $O_1(\rs)$ &           5.5 &              9.1 &              12.7              \\
    $O_2(\rs)$ &           0.5 &              0.9 &              1.3               \\
    MAX$_{1}(\rs)$ &       10 &               12&                16                \\
    MAX$_{2}(\rs)$ &       2&                 3&                 4                 \\
    QGS($0.2\mu\text{s}$)  &                  $30\cdot 10^{6}$ & $50\cdot 10^{6}$ & $70\cdot 10^{6}$\\
    QGS($2\mu\text{s}$) &   $3\cdot 10^{6}$ &  $5\cdot 10^{6}$ &  $7\cdot 10^{6}$\\
    \bottomrule
  \end{tabular}
  \end{center}
\end{table}

Figure~\ref{fig:circuitdepth_GHZ}c)
compares the circuit depth as a function of the distance $\rg$ for different
system sizes. The truncation of states during the
search does not necessarily converge to the global minima.
%
%
The compiler settings allow one to truncate up to fifty percent of the configurations
if they have two or more qubits less in the GHZ state than the best configuration.
Afterward, we keep 300 geometries with the best metrics.
An example of the $4 \times 4$ grid with minimal distance $\rg = 2 \al$
is shown for the hexagonal lattice layout in Fig.~\ref{fig:circuitdepth_GHZ}b).
The possible reduction of the circuit depth increases with the system size.
For the circuits of Figs.~\ref{fig:setup}, \ref{fig:effects_rydberg}, and
\ref{fig:meas_rydberg}, we keep up to 600 geometries during compiling the
algorithm.

The front end of the tensor network software takes care of scheduling the parallel
pulses according to the compiler output, including the distance~$\rg$. The scheduling includes here the individual addressing of the atoms, i.e., activating the pulses on the selected atoms and setting no laser-driving terms for atoms not involved in any gate. The pulses for the Rabi frequency $\Omega^{R}$
driving the transition $\ket{1} \leftrightarrow \ket{r}$ and the corresponding detuning
are specified in Ref.~\cite{Pagano2022}, where the detuning follows a Gaussian shape
and the Rabi frequency is $10\,\text{MHz}$. The total duration of a single
CZ is $0.122\,\mu\text{s}$ and we choose the amplitudes for the single-qubit
gates to match this duration of the CZ gate. The TTN simulation uses this a duration of $0.122\,\mu\text{s}$
independent from the scenarios for decoherence, i.e., the theoretical estimate for decoherence according to Eq.~\eqref{eq:dephasing} with
$\tlayer = 0.2\, \mu\text{s}$ and $\tlayer = 2\mu\text{s}$ is calculated in a post-processing step without accessing TTN results. However, the
time $\tlayer$ becomes a relevant variable of the platform as soon as we
consider decoherence, recall Fig.~\ref{fig:meas_rydberg}d). We
report selected key statistics of the circuits in Tab.~\ref{tab:statistics}. Especially
the $8 \times 8$ grid demonstrates how larger system sizes profit from the
parallelization with $35\%$ speedup in comparison to a circuit serial in CZ gates
as well as $92\%$ speedup compared to a completely serial circuit. In contrast,
we are only $15\%$ above the theoretical minimum shown in Tab.~\ref{tab:statistics}.
The maximal number of local gates and CZ gates further gives an
estimate of which laser power is required to execute gates in parallel. For
cross-platform comparisons, the level of parallelization is only one aspect,
but the gates executed per time unit are more meaningful. Hence, we also list the
number of Quantum Gates per Second (QGS) for the two scenarios, e.g.,
the preparation of the global GHZ state on an $8 \times 8$ grid in the more
conservative scenario still runs approximately $7 \cdot 10^{6}$ gates per second.

\section{Conclusion}                                                            \label{sec:concl}
%
We have analyzed the parallel execution of a quantum algorithm on the digital
twin of a quantum computer. The simulation of an algorithm on the pulse
level involving the Hamiltonian leads to valuable insights into the quantum
hardware, e.g., in terms of qubit crosstalk. 
In detail, the numerical simulations analyze the
influence of the Rydberg interactions on the parallel execution of
controlled-phase gates. We have simulated
the qutrits physics in systems of up to a size of $8 \times 8$ sites with a tree tensor
network. The simulations
cover a variety of effects, including decay effects modeled by a Lindblad channel
and the remaining population in the Rydberg state, where the latter would
require non-Markovian effects in a two-level open system description.

The focus on the preparation of a GHZ state explores one of the first standard demonstrations of control and stability of new quantum processors. 
We have shown that the circuit depth for the preparation of a GHZ state
with almost 700 gates for an $8 \times 8$ grid can be reduced by
more than 35\% percent under the constraint
of avoiding parallel gates for a radius smaller than $\rs = \sqrt{16} \al$.
Satisfying this condition, the infidelity
per single CZ gate in the closed Rydberg quantum
system increases by one or two orders of magnitude in comparison to the infinite-$\rg$ limit. However,  
the final infidelity stays a factor of $5$ to $10$ below the major source of errors introduced by the decay from the Rydberg state.
Furthermore, we have shown that the
infidelity due to decay from the Rydberg state stays constant beyond
a certain circuit depth. Dephasing, i.e., the next most important open quantum
system effect, represents the incentive to minimize the circuit depth
as long as the crosstalk is under control. Due to the speedup of $92\%$ and
the inherent reduction in decoherence revealed by the digital
twin in comparison to a completely serial circuit, we identify quantum gate parallelism
as a necessary pathway to maximize the performance of NISQ devices.

Building on this initial investigation, we foresee a number of important questions that now can and shall be addressed to support future NISQ development, along the lines of the three challenges for future quantum hardware outlined in the white paper~\cite{Wack2021}. For example, 
the development -- also via optimal control methods~\cite{Mueller2022,Koch2022} -- of a four-qubit gate performing two CZ gates with suppression of crosstalk could highly improve the system performance for Rydberg systems. 
We foresee an engineering challenge to design the embedded classical hardware
to control and apply the laser pulses in parallel.
Questions related to the green quantum advantage and how parallel execution can
be leveraged to shift the boundary for energy-efficient computing
more toward QPUs shall be addressed~\cite{Jaschke2022}.
Moreover, Rydberg platforms offer two pathways to modify the connectivity. On the
one hand, the Rydberg blockade radius can be modified which affects the range at
which two-qubit gates are possible~\cite{Spierings2022}. On the other hand, shift operations offer
another degree of freedom: the Rydberg atoms trapped
in the optical tweezers can be shifted during an algorithm as demonstrated
in~\cite{Bluvstein2021}. This additional degree
of freedom can possibly be exploited to further improve algorithm efficiency.

The digital twin introduced here can be straightforwardly applied to different quantum computing hardware, such as superconducting or ion platforms.
Several questions remain open on the theory side, such as gaining a better understanding of the running details of other algorithms and platforms.
For example, a detailed comparison of the highly
parallelized GHZ state preparation in Ref.~\cite{Piroli2021} is another intriguing
direction, where the protocol uses auxiliary qubits and measurements.
As is, our implementation of a digital twin is directly relevant for NISQ
devices with the current number of simulated qubits.
Different improvements on the software and hardware aspects of the digital twin, e.g., via parallelization on classical computers~\cite{Secular2020}, will enable in the near future exploring higher entangling algorithms and larger system sizes. 

\emph{Acknowledgments $-$}
We thank Marco Ballarin, Timo Felser, and Marco Trenti for their work on the
underlying numerical library \emph{Quantum TEA}; our partners from the QRydDemo project Hans Peter B{\"u}chler,
Florian Meinert, Tilman Pfau, and J{\"u}rgen Stuhler for useful discussions; Marcello Dalmonte, Nora Reini{\'c}, and Ilaria Siloi for feedback.
This project has received funding from the German Federal Ministry of Education and Research (BMBF) under the funding program quantum
technologies $-$ from basic research to market $-$ with the grant QRydDemo.
We acknowledge support from the Italian Ministry of University and Research (MUR) via PRIN2022-TANQU and the Departments of Excellence grant 2023-2027 Quantum Frontiers; we acknowledge support from the European Union via H2020 projects Pasquans2, EURyQA and
TEXTAROSSA, the EU-QuantERA projects QuantHEP and T-NISQ, the WCRI-Quantum Computing
and Simulation Center of Padova University, and the Italian National Centre on HPC,
Big Data and Quantum Computing.
This work was performed in part at the Aspen Center for Physics, which is supported by National Science Foundation grant PHY-2210452; the participation of D. J. at the Aspen Center for Physics was supported by the Simons Foundation.
The authors acknowledge support by the state of Baden-W{\"u}rttemberg through
bwHPC and the German Research Foundation (DFG) through grant no
INST 40/575-1 FUGG (JUSTUS 2 cluster).

\bibliography{bibliography/refs,bibliography/specific_refs}

\begin{thebibliography}{75}
\bibitem [{QTN()}]{QTNoJumps}%
  \BibitemOpen
  \href@noop {} {\enquote {\bibinfo {title} {{The first step of quantum
  trajectories relies on the non-hermitian evolution of the quantum system
  which causes a drop in the norm. For each trajectory, a random number $r$ is
  drawn and a jump is applied as soon as the norm drops below $r$. In our case,
  we are interested in the probability of such an event, but we are not
  considering recovering decaying states after the jump. Thus, we have no
  additional benefit of averaging over trajectories which all lead to a zero
  fidelity with the GHZ states as they have at least one qubit in the state
  $\ket{d}$ outside the computational basis of the qubit. In summary, the lost
  norm is sufficient to obtain the percentage of incorrect preparations.}}}\
  }\BibitemShut {NoStop}%
\bibitem [{Dmi()}]{DminBound}%
  \BibitemOpen
  \href@noop {} {\enquote {\bibinfo {title} {{The theoretical bound for the
  circuit depth assumes CZ running gates on nearest-neighbors, i.e., the
  control qubit of the first CZ gate can have its target qubit as left
  neighbor, while its right neighbor is the control qubit of a second CZ gate.
  This approach leads unavoidable to crosstalk as there are no means for this
  control qubit to distinguish which nearest-neighbor is the target qubit in
  its CZ gate. As the CZ gate is symmetric for the control and target qubit,
  the case of having the target qubit of another CZ as a nearest-neighbor leads
  to the same error.}}}\ }\BibitemShut {NoStop}%
\end{thebibliography}

\appendix

\section{Algorithms for compiling and scheduling GHZ states}                    \label{app:compiler}
%
We require a compiler minimizing the final
circuit depth for the given constraint of crosstalk induced by an interaction
decaying with the distance between the qubits. The minimization of the depth $-$ not
the number of gates $-$ and the specific constraints are up to our knowledge not easily
accessible in current software solutions. We encounter the problem to
account for a)~a figure of merit that minimizes the circuit depth of a
parallel circuit, and b)~blocking atoms around an atom executing a CZ gate
for a given radius.

The focus of compilers is so far on minimizing the number of gates, which
minimizes the circuit depth of a serial circuit; for example, implementations like SABRE tackle this problem~\cite{Li2019}. To generate an $N$-qubit
GHZ state, we need exactly $N - 1$ entangling gates, which is already the optimum.
We require a way to minimize the depth of a parallel circuit.
Thus, one option to make this way work is to define multi-qubit gates as native
gates that encode two, three, or even more CZ gates. For example, the native
gate encoding two CZ gates is $\mathrm{CZCZ}_{ijkl} = \mathrm{CZ}_{ij} \mathrm{CZ}_{kl}$ where
$i,j,k,l$ encode the position of the atoms. Given the assumption that a
compiler can translate between native gates given the corresponding
rules, we have to define a large number of gates for larger systems with $N$ atoms,
e.g., $N (N - 1) / 2$ starting with the pairs of CZ gates. These large
number of gates likely lead to an unfavorable scaling when optimizing
the problem.

The second option is then to go via the CZ gate as a native
gate and use the connectivity given by compilers.
The connectivity or coupling map of a QPU informs the compiler about which pairs
of qubits can be used in an entangling gate without swap operations. Before adding
the first CZ gate to a layer, the map starts with all nearest-neighbor connections;
After adding a CZ gate, the map would have to be updated dynamically during compilation
for building each parallel layer. Another possible option is to pass a set of possible
coupling maps that can be used for each parallel layer, but here we face a large
set of possibilities that does not scale well with system size.

\begin{figure}[t]
  \begin{center}
    \begin{minipage}{0.99\linewidth}
      \begin{overpic}[width=1.0 \columnwidth,unit=1mm]{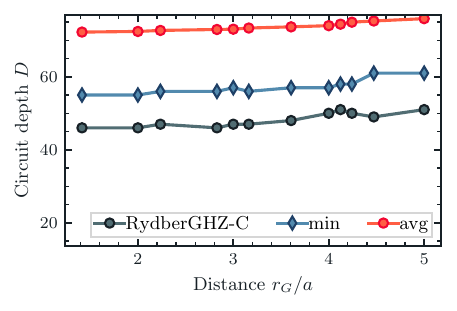}
        \put( 0,63){a)}
      \end{overpic}
    \end{minipage}\hfill
    \begin{minipage}{0.99\linewidth}
      \begin{overpic}[width=1.0 \columnwidth,unit=1mm]{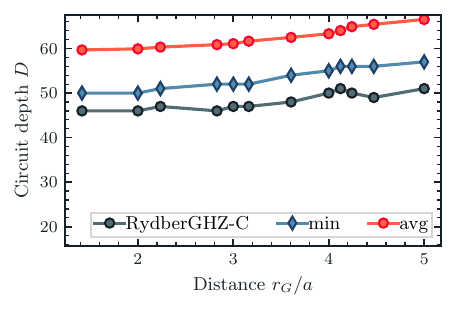}
        \put(0,63){b)}
        \put(80, 2){\figsharesapp{}}
      \end{overpic}
    \end{minipage}
    \caption{\emph{Benchmarking the RydberGHZ-C against random configurations for $8 \times 8$ square lattice.}
      We compare the results of the RydberGHZ-C against the average (avg) circuit depth
      of 1000 randomized realizations of the circuit and the minimum circuit depth
      out of the random set.
      a)~Random sequence of CNOT gates with the initial Hadamard gate placed
      in the middle of the circuit.
      b)~Random sequence of CNOT gates with the initial Hadamard gate placed
      in the middle of the circuit and a waiting period of four cycles between
      CNOT gates. The waiting period is useful for accounting for the four local
      gates which have to run in between two CZ gates with the given native gate
      set.
                                                                                \label{fig:cbenchmark}}
  \end{center}
\end{figure}

Each of the above modifications for the compiler require a significant modification
of existing approaches.
For these reasons, we aim for an algorithm which can minimize the circuit
depth of the parallel GHZ state preparation on the Rydberg platform and serves
as a compiler for our simulations. Figure~\ref{fig:cbenchmark} serves as a
benchmark that the compilation is meaningful, where we consider two scenarios for an $8 \times 8$ square lattice.
First, we consider a circuit with the initial Hadamard gate in the center of
the square lattice and then draw 1000 random sequences of the CNOT gates, see
Fig.~\ref{fig:cbenchmark}a). An important step in the compiler algorithm is
the consideration of four local gates which are placed in between two CZ
from the perspective of a single qubit; the condition can also be easily implemented here:
if other qubits are available for an entangling gate, qubits are blocked for four
cycles after each entangling gate and we draw the next step from the remaining
possible options. This rule significantly improves the circuit depth of the best
out of 1000 random sequences as shown in Fig.~\ref{fig:cbenchmark}b), but the actual
compiler still has a small advantage. Especially the second comparison which
already has two key features of the optimal solution encoded, i.e., the central
starting point and the consideration of the four local gates, confirms that
the compilation is meaningful and we explain the approach now in more detail.

\begin{figure}[t]
  \begin{center}
    \begin{minipage}{0.99\linewidth}
      \begin{overpic}[width=1.0 \columnwidth,unit=1mm]{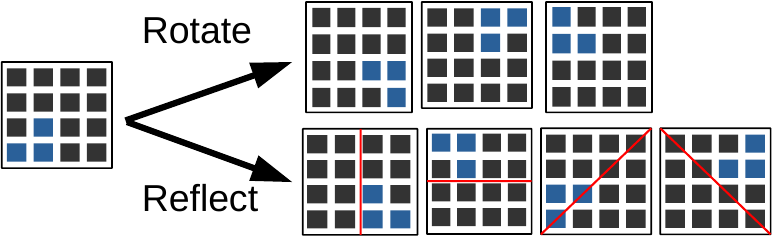}
        \put( 0,25){a)}
      \end{overpic}
    \end{minipage}\hfill
    \vspace{0.2cm}
    \begin{minipage}{0.99\linewidth}
      \begin{overpic}[width=1.0 \columnwidth,unit=1mm]{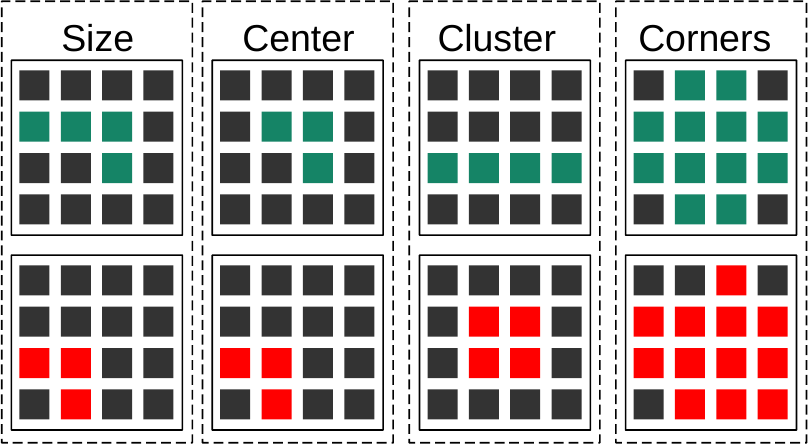}
        \put(0,57){b)}
        \put(80, -4){\figsharesapp{}}
      \end{overpic}
      \vspace{0.15cm}
    \end{minipage}
    \caption{\emph{Building blocks for compiling the algorithms.}
      a)~Symmetries lead to configurations which are equal by
        rotations or reflections; only one of them has to be kept
        for further steps.
      b)~Truncation rules assign priority to more promising
        configurations (shown in green) over less promising
        configurations (shown in red). The rules consider the
        current size in the intermediate GHZ state, how close
        the center of mass is to the center, if the current
        sites form a cluster where all sites are within a
        small radius, and the size of the remaining
        corners.
                                                                                \label{fig:copmiler}}
  \end{center}
\end{figure}

We recall the compiler steps from Sec.~\ref{sec:rydberghz}:

\emph{
\begin{enumerate}[noitemsep,nolistsep]
  \item{Fix minimal distance $\rg$, i.e., the minimal
    distance between any two atoms participating in different CZ
    gates at the same time.}
  \item{A set of unique sites is selected as initial starting points for the
     Hadamard gate and store them as a list of configurations
     $\mathcal{C}_{\gamma}$. No unique site inside the set maps via
     rotations or reflections of the grid to another site within the set, see Fig.~\ref{fig:copmiler}a).}
  \item{List all possible pairs $\mathcal{P}_{\alpha}$ of control and target sites for the
    configurations $\mathcal{C}_{\gamma}$.}
  \item{List all sets $\mathcal{S}_{\beta}$ of pairs $\mathcal{P}_{\alpha}$ which
    can run in parallel while obeying the minimal distance $\rg$.
   Subsets are not included, e.g., if 
   $\mathcal{S}_{\beta} = (\mathcal{P}_{A}, \mathcal{P}_{B}, \mathcal{P}_{D})$
   is included, we omit $\mathcal{S}_{\beta'} = (\mathcal{P}_{A}, \mathcal{P}_{B})$.}
  \item{Create a new configuration $\mathcal{C}_{\gamma}'$ for each $\mathcal{S}_{\beta}$. Check
    if the new configuration $\mathcal{C}_{\gamma}'$ is a solution.}
  \item{Block the qubits involved in a CZ gate for the next four iterations
    to allow the application of the local gates for the Rydberg platform.
    Reduce iterations blocked for local operations by one for all qubits
    not involved in a CZ gate.}
  \item{Continue with step 3) using the new configurations
    $\mathcal{C}_{\gamma}'$ as $\mathcal{C}_{\gamma}$. Allow to truncate
    entries in $\mathcal{C}_{\gamma}$ based on symmetry arguments
    or truncation rules, see Fig.~\ref{fig:copmiler}.}
\end{enumerate}
}

The second step and the comment regarding the rotations and reflections are
shown in detail in Fig.~\ref{fig:copmiler}a): for the example of a $4 \times 4$
grid, we show the three rotations and four reflections considered during the
compilation to eliminate redundant solutions. The illustration shows the
three relevant sites for the initial Hadamard gate in blue and how these
three sites cover all configurations with the rotations and reflections. The
equal approach can be carried out at later stages to find unique sets of
configurations.

Figure~\ref{fig:copmiler}b) sketches the rules to truncate configurations
in step 7). The selection is implemented
following four conditions. The first condition simply counts the number of
qubits already added to an intermediate GHZ state after a given number of
parallel layers, where we favor configurations that already have more qubits
in the intermediate GHZ states, see label "Size". As mentioned before, the initial
Hadamard gate close to the center is favorable for reaching all corners in
the system in an equal number of steps; therefore, we generalize this premise
and calculate the center of mass of the intermediate GHZ state and prefer
states where the center of mass is close to the center of the grid ("Center").
The next condition is more subtle and concerns the chance to run CZ gates
in parallel in the next layers: configurations arranged in a cluster likely
fit in one or a few circles of radius $\rg$, while widely spread configurations
cannot be easily covered by a few circles of the same radius. Thus, a higher
standard deviation or mean distance from the center of mass of a configuration
is beneficial for scheduling the next set of parallel CZ gates, see "Cluster". The fourth
condition is the inverse of the third and avoids that there remains a large
corner outside the GHZ state which cannot be finalized efficiently at the end,
i.e., the "Corner" rule. Together, these four rules on the size, center, cluster,
and corners decide on which configurations are kept for the next iteration.
With many configurations truncated, this truncation leads to noise in the
circuit depth as seen in Fig.~\ref{fig:effects_rydberg}b): even with
a larger distance $\rg$ enforced, we find a lower circuit depth than for
smaller distances $\rg$.

\end{document}